\documentclass[12pt,showpacs,pra]{revtex4}
\usepackage{graphics,graphicx}
\usepackage{dcolumn}
\usepackage{bm}
\usepackage{amsfonts}
\usepackage{amssymb}
\usepackage{amsmath}
\usepackage{txfonts}

\newcommand{\Tr}        {\mathrm{Tr}}

\newcommand{\bra}[1]    {\left\langle #1 \right|}
\newcommand{\ket}[1]    {\left| #1 \right\rangle}

\begin{document}
\title{Purifying GHZ States Using Degenerate Quantum Codes}
\author{K. H. Ho and H. F. Chau}
\date{\today}
\affiliation{Department of Physics and Center of Theoretical and Computational
Physics, The University of Hong Kong, Pokfulam Road, Hong Kong, China}
\begin{abstract}
Degenerate quantum codes are codes that do not reveal the complete error
syndrome.  Their ability to conceal the complete error syndrome makes them
powerful resources in certain quantum information processing tasks.  In
particular, the most error-tolerant way to purify depolarized Bell states using
one-way communication known to date involves degenerate quantum codes.  Here we
study three closely related purification schemes for depolarized GHZ states
shared among $m \geq 3$ players by means of degenerate quantum codes and
one-way classical communications.  We find that our schemes tolerate more noise
than all other one-way schemes known to date, further demonstrating the
effectiveness of degenerate quantum codes in quantum information processing.  
\end{abstract}
\pacs{03.67.Hk, 03.67.Pp, 89.70.Kn}
\maketitle
\section{Introduction}
Quantum error correcting codes, unlike their classical counterparts, may not
reveal the complete error syndrome.  Codes with this property are known as
degenerate codes~\cite{SS96,DSS98}.
In a sense, degenerate codes pack more information than non-degenerate ones
because different quantum errors may not take the code space to orthogonal
spaces.  By carefully utilizing the degenerate property, degenerate codes are
useful resources in quantum information processing.  Examples showing their
usefulness were provided by Shor and his co-workers~\cite{SS96,DSS98}.  In
particular, they showed that a
carefully constructed degenerate code is able to purify Bell states passing
through a depolarizing channel with fidelity greater than
0.80944~\cite{DSS98}.  Their scheme is more error-tolerant than all the
known one-way depolarized Bell state purification schemes involving
non-degenerate codes to date.

It is instructive to ask if the degenerate codes can be used to improve the
error-tolerant level of existing one-way multipartite purification protocols.
Here we provide such an example by considering the purification of shared
GHZ states.  Specifically, suppose that a player prepares many copies of
perfect GHZ state in the form
\begin{equation}
 \ket{\Phi^{m+}} \equiv \frac{1}{\sqrt{2}} \left( \ket{0^{\otimes m}} +
\ket{1^{\otimes m}} \right) ~.
 \label{eq:GHZ}
\end{equation}
For each perfect GHZ state, he/she keeps one of the qubit and sends the other
to the remaining players through a depolarizing channel so that upon reception
of their qubits, these $m$ players share copies of Werner state
\begin{equation}
 W_F = F \ket{\Phi^{m+}} \bra{\Phi^{m+}} + \frac{1-F}{2^m - 1}
 \left( I - \ket{\Phi^{m+}} \bra{\Phi^{m+}} \right) ~,
 \label{eq:Wf}
\end{equation}
where $F$ is the fidelity of the channel and $I$ is the identity operator.
Now, the players wanted to distill shared perfect GHZ states using an one-way
purification scheme that works for as small a channel fidelity as possible.
Clearly, this task is a generalization of the Bell state distillation problem
investigated by Shor and his co-workers~\cite{SS96,DSS98}.

We begin our study by defining a few notations and reviewing prior arts in
Sec.~\ref{sec:GHZ_re}.  Then we introduce three closely related one-way
multipartite purification protocols involving concatenated degenerate codes and
analyze their performances in Sec.~\ref{sec:our_pro}.  Actually, all three
protocols use the same repetition code as their inner codes.
Moreover, in the case of $m = 2$, one of the our protocols is a generalization
of the scheme proposed by DiVincenzo \emph{et al.}~\cite{DSS98}.  Most
importantly, for $m\geq 3$, our protocols are the most error tolerant ones
discovered so far in the sense that ours can distill shared GHZ states from
copies of Werner state in
the form of Eq.~(\ref{eq:Wf}) with a fidelity $F$ so low that no other one-way
purification schemes known to date can.
Our schemes can also be generalized to the case when the Hilbert space
dimension of each quantum particle is greater than 2.  We briefly discuss this
issue in Sec.~\ref{sec:high_spin}.  Finally, we summarize our findings in
Sec.~\ref{sec:sum}.

\section{Prior Arts}
\label{sec:GHZ_re}
\subsection{Some notations}
Given that $m\geq 2$ players share $N$ noisy GHZ states in the form of
Eq.~(\ref{eq:GHZ}).  Clearly, the GHZ state is stabilized by its stabilizer
generators, namely,
\begin{eqnarray} \label{eq:GHZ_gen}
S_0 &=& X_0 X_1 \cdots X_{m-1} ~, \nonumber \\
S_i &=& Z_0 Z_i 
\end{eqnarray}
for $1\leq i \leq m - 1$, where
\begin{equation}
X_i = \left( \begin{array}{cc} 0 & 1 \\ 1 & 0 \end{array} \right) ~,
Z_i = \left( \begin{array}{cc} 1 & 0 \\ 0 & -1 \end{array} \right)
\end{equation}
denote the spin flip and phase shift operation acting on the $i$th qubit
respectively.
For simplicity, we use the shorthand notation 
$(\beta, \boldsymbol{\alpha}) \equiv
(\beta, \alpha_1, \alpha_2, \ldots, \alpha_{m-1}) \in GF(2) \times GF(2)^{m-1}$
to denote the eigenvalues of stabilizer generators.
Here $\beta \in GF(2)$ is the eigenvalue
of the operator $S_0$, namely, the phase error detected; and $\alpha_i \in
GF(2)$ is the eigenvalue of the operator $S_i$, namely, the bit flip error
detected, for $1\leq i\leq m - 1$.  We sometimes abuse the notation
to denote a state by $(\beta, \boldsymbol{\alpha})$.
That is, we denote the states $(|0^{\otimes m}\rangle +
|1^{\otimes m}\rangle)/\sqrt{2}$ and $(|0^{\otimes m}\rangle -
|1^{\otimes m}\rangle)/\sqrt{2}$ by $(\beta, \boldsymbol{\alpha}) = (0,
\boldsymbol{0})$ and $(\beta, \boldsymbol{\alpha}) = (1, \boldsymbol{0})$
respectively.

\subsection{Depolarization to the GHZ-basis diagonal states}
The $m$ players can depolarize each copy of their shared GHZ state into the GHZ
diagonal basis using local operation and classical communication (LOCC) in the
following way~\cite{DCT99}. 
A player randomly chooses an operator from the span of the set of stabilizer
generators of the GHZ state and broadcast his/her choice to the other players.
Then they collectively apply the chosen operator to the GHZ state.  Since all
stabilizer generators of the GHZ state in Eq.~(\ref{eq:GHZ_gen}) are tensor
products of local unitary operators $X_i$ or $Z_i$, the players can apply the
operator chosen above to the state locally using LOCC.  Then, they forget
which operator they have chosen.  The resultant state is diagonal in the GHZ
basis.  Moreover, the error rate of the GHZ state is unchanged in this process.
So, we can always assume that each state shared among the players are
diagonal in the GHZ basis.

Among all GHZ-basis diagonal states with a fixed error rate (and hence also
among all states with a fixed error rate), Werner state is the most difficult
to work on as far as distillation of GHZ states is concerned.  This is because
one can turn any state into a Werner state with the same error rate via a
depolarizing channel.  Hence, to study the worst case performance of the
distillation of GHZ states, we suffices to investigate the case in which the
input states are Werner states.

\subsection{Maneva and Smolin's multi-party hashing protocol and its
generalization by Chen and Lo}
\label{sec:MH_pro}
Maneva and Smolin~\cite{MS00} proposed a multi-party hashing protocol by
generalizing the bilateral quantum XOR (BXOR) operation~\cite{BDSW96a} to the
multipartite case.  In their scheme, the $m$ players
carefully choose two (classical) random hashing codes, one to
correct spin flip errors and the other to correct phase errors, and apply them
to their shared noisy GHZ states.  This can be done by using local operation
plus classical communication with the help of a few multi-lateral
quantum XOR (MXOR) operations.
Recall that MXOR is a linear map transforming the state
$\bigotimes_{i=0}^{m-1} |j_i,k_i\rangle_i$ to $\bigotimes_{i=0}^{m-1}
|j_i,j_i+k_i\rangle_i$ for all $j_i,k_i\in GF(2)$.  Here, quantum particles
with subscript $i$ belong to the $i$th player.
Suppose the source and target states are eigenstates of the stabilizer
generator in Eq.~(\ref{eq:GHZ_gen}) with eigenvalues $(\beta_1,
\boldsymbol{\alpha}_1)$ and $(\beta_2, \boldsymbol{\alpha}_2)$ respectively.
Then after the MXOR operation, the resultant state is also an
eigenstate of the stabilizer generator with eigenvalues~\cite{MS00}
\begin{equation}\label{eq:MXOR_Def}
\text{MXOR}[(\beta_1, \boldsymbol{\alpha}_1),
(\beta_2, \boldsymbol{\alpha}_2)] =[(\beta_1 + \beta_2, \boldsymbol{\alpha}_1),
(\beta_2, \boldsymbol{\alpha}_1 - \boldsymbol{\alpha}_2)] ~.
\end{equation}

Using the observation that spin flip error occurred in different qubit
of a GHZ state can be detected and corrected in parallel,
Maneva and Smolin showed that the asymptotic yield of their hashing protocol 
in the limit of large number of shared noisy GHZ states is given by~\cite{MS00}
\begin{equation}\label{eq:D1}
D_1 = 1 - \max_{1 \leq i \leq m - 1 }[{H(b_i)}] - H(b_0) ~,
\end{equation}
where $H(x) \equiv - \sum_j p_j \log_2 p_j$ is the classical Shannon entropy 
function. Here the $n$-bit string $b_0$ represents the random choice
of $\beta_1, \ldots, \beta_N$ where $\beta_\ell$ corresponds to the eigenvalue 
of the operator $S_0$ of the $\ell$th GHZ-state $\ket{\Phi^{m+}}$'s and
the $N$-bit string $b_i$ represents the random choice
of $\alpha_{1i}, \ldots, \alpha_{Ni}$ where $\alpha_{\ell i}$ is the eigenvalue
of the operator $S_i$ of the $\ell$th GHZ state for $1\leq i\leq m - 1$.
That is to say, $H(b_0)$ is the averaged phase error rate
and $H(b_i)$ is the averaged bit flip rate corresponding to the stabilizer
generator $S_i$ (for $i=1,2,\ldots ,m-1$) over the $N$ GHZ states respectively.

Recently, Chen and Lo improved the above random hashing protocol by exploiting
the correlation between the string $b_i$.  Specifically, they replaced the
spin flip error-correction random hashing code used in
Maneva and Smolin protocol by the following scheme.  (Actually, they only
considered the case of three players.  What we report below is a
straight-forward generalization to the case of $m$ players as we need to
use this generalization later on.) Player~1
applies (classical) random hashing to correct spin flip error occurred in his/her
share of the GHZ states.  He/She then broadcasts his/her hashing code used and
his measurement results.  For $i = 2,\ldots , m-1$, the $i$th player
carefully picks his/her (classical) random hashing code to correct spin flip error occurred
in his/her share of the GHZ states
based on the broadcast information of players~$1,2,\ldots , i-1$.  Then,
the $i$th player broadcasts his/her code used and his/her measurement results.
In this way, the yield of Maneva and Smolin scheme can be
increased to~\cite{CL07}
\begin{eqnarray}\label{eq:D2}
D_2 & = & 1 - \max \{ H(b_1), H(b_2 | b_1), H(b_3 | b_2, b_1), \cdots
, H(b_{m-1} | b_{m-2},b_{m-3}, \ldots , b_1) \} \nonumber \\
& & ~- H(b_0) + I(b_0; b_{m-1}, b_{m-2}, \ldots , b_1)
\end{eqnarray}
where the function $I(\,;\,)$ is the mutual information between the two
classical random variables appear in its arguments.

Applying the random hashing method of Maneva and Smolin to a collection of
identical tripartite (that is, $m = 3$) Werner states in Eq.~(\ref{eq:Wf}), one
can obtain perfect GHZ state with non-zero yield whenever the fidelity
$F\geq 0.8075$~\cite{MS00}.  Using the Chen and Lo's formula in
Eq.~(\ref{eq:D2}), one can push this threshold fidelity down to
$0.7554$~\cite{CL07}.

\subsection{Shor-Smolin concatenation procedure and its generalization to
 the multipartite situation}
\label{sec:SS_pro}
Built on an earlier work by Shor and Smolin~\cite{SS96},
DiVincenzo \emph{et al.} introduced a highly error-tolerant way of distilling
shared Bell states by means of a concatenation procedure~\cite{DSS98}.
This procedure can be generalized to distill shared GHZ states in a
straight-forward manner.  We report this generalization below since we have to
use a few related equations later on.

Suppose $m$ players share $N n$ copies of imperfect GHZ states for $N\gg 1$.
They perform the following two level decoding procedure.
First, the players randomly divide these GHZ states into $N$ equal parts.
Then each player applies a
decoding transformation associated with an additive $[n,k_1,d_1]$ code
to his/her own qubits followed by the error syndrome measurements.  Surely,
this can be done with the help of a few MXOR operations.
By comparing the difference 
in player's measurement results, they obtain the syndrome $\vec{\boldsymbol{s}}
\in GF(2)^{(m-1)(n-k_1)}$.
To continue, each party applies another decoding transformation corresponding to a
(classical) random hashing
code $[N k_1, k_2, d_2]$ to correct errors in the GHZ diagonal basis and broadcast the measurement results. Finally, they 
apply the necessary unitary transformation according to the measured error
syndrome of this random hashing code to get the purified GHZ states.

Suppose that an additive code $[n, k_1, d_1]$ is applied and the remaining
states after the decoding 
transformation and measurements are denoted by $(\delta,\boldsymbol{\gamma})
\equiv \text{TRAN}[(\beta_1, \boldsymbol{\alpha}_1),
(\beta_2,\boldsymbol{\alpha}_2),
\ldots, (\beta_{k_1}, \boldsymbol{\alpha}_{k_1})]$.  Then,
the yield of this concatenated scheme is given by the so-called Shor-Smolin
capacity~\cite{SS96,DSS98}
\begin{equation} \label{eq:D_SS}
D_\textrm{SS} = \frac{1}{n} ( 1 - S_X) ~,
\end{equation}
where
\begin{equation} \label{eq:Sx} 
S_X = \sum_{\vec{\boldsymbol{s}} \in GF(2)^{(m - 1) ( n - k_1)}} \text{Pr}
(\vec{\boldsymbol{s}}) \enspace h(\{\text{Pr} 
((\delta, \boldsymbol{\gamma}) | \vec{\boldsymbol{s}}) : (\delta,
\boldsymbol{\gamma}) \in GF(2)^m \})
\end{equation}
is the average of the von Neumann entropies of the quantum states conditional
on the measurement outcomes.  Note that in Eq.~(\ref{eq:Sx}),
$\text{Pr} (\vec{\boldsymbol{s}})$ is the probability that the measurement
outcome is $\vec{\boldsymbol{s}}$,
\begin{equation}
 h(\{ p_i \} ) \equiv - \sum_i p_i \log_2 p_i ~,
\end{equation}
and
\begin{equation}
 \sum_i p_i = 1 ~.
\end{equation}

By applying the above procedure to depolarized Bell states
using a 5-qubit cat code as the inner code and a random hashing code as the
outer code (that is, the case of $m = 2$ and $n = 5$),
DiVincenzo \emph{et al.} found that one can attain a non-zero capacity whenever
the channel fidelity $F > 0.80944$~\cite{DSS98}.  Since the performance of
this scheme exceeds that of quantum random hashing code and that
the 5-qubit cat code is degenerate, the power of using degenerate quantum code
in quantum information processing is demonstrated.

\subsection{Other hashing and breeding schemes}
Several other multipartite hashing schemes have been studied~\cite{MS00,HDM06}.
In particular, Maneva and Smolin's hashing scheme can distill shared GHZ states
from copies of Werner states with fidelity $F \geq 0.7798$ in the limit of
arbitrarily large number of players (that is, when $m \rightarrow
\infty$)~\cite{MS00}.
Another approach is to use the so-called stabilizer breeding.  In particular,
Hostens \emph{et al.} showed that stabilizer breeding is able to purify
depolarized $5$-qubit ring state with fidelity
$F\geq 0.756$~\cite{HDM06b}.  A few authors also studied the distillation of
graph state subjected to local $Z$-noise~\cite{KPDB06} and
bicolorable graph state.~\cite{GMR06}  Furthermore, Glancy
\emph{et al.} generalized the hashing protocol of Maneva and Smolin to
purify a much larger class of output state.~\cite{GKV06}
The second column in Table~\ref{tb:pw} summarizes
the state-of-the-art one-way purification schemes to distill depolarized GHZ
states before our work.

\begin{center}
\begin{table}
\begin{tabular}{||c|c|c|c||} 
\hline
\hline
~~$m$~~ & prior art & our best protocol & lower bound \\
\hline
2 & 0.8094 & 0.8094 & 0.7500 \\
3 & 0.7554 & 0.7074 & 0.6111 \\
4 & 0.7917 & 0.6601 & 0.5500 \\
\hline
\hline
\end{tabular}
\caption{The threshold fidelity of the depolarizing channel above which a GHZ
state can be distilled by prior art and by our protocols.  Also listed is the
lower bound of the fidelity below which no one-way protocol can distill shared
GHZ state using Eq.~(\ref{eq:F_min_lower_bound}) in Sec.~\ref{subsec:limit}.
As for prior art, the threshold fidelity for $m = 2$ is given by the 5-qubit
cat code~\cite{DSS98}.  For $m = 3$ case, the threshold is computed by the Chen
and Lo's formula~\cite{CL07} in Eq.~(\ref{eq:D2}).  For $m = 4$, the threshold
is given by the Maneva and Smolin's hashing protocol~\cite{MS00} in
Eq.~(\ref{eq:D1}).
\label{tb:pw}
}
\end{table}
\end{center}

\section{Our protocols involving degenerate code and their performances}
\label{sec:our_pro}

\subsection{Our protocols}

\begin{figure}
\centering\includegraphics[width=14.5cm]{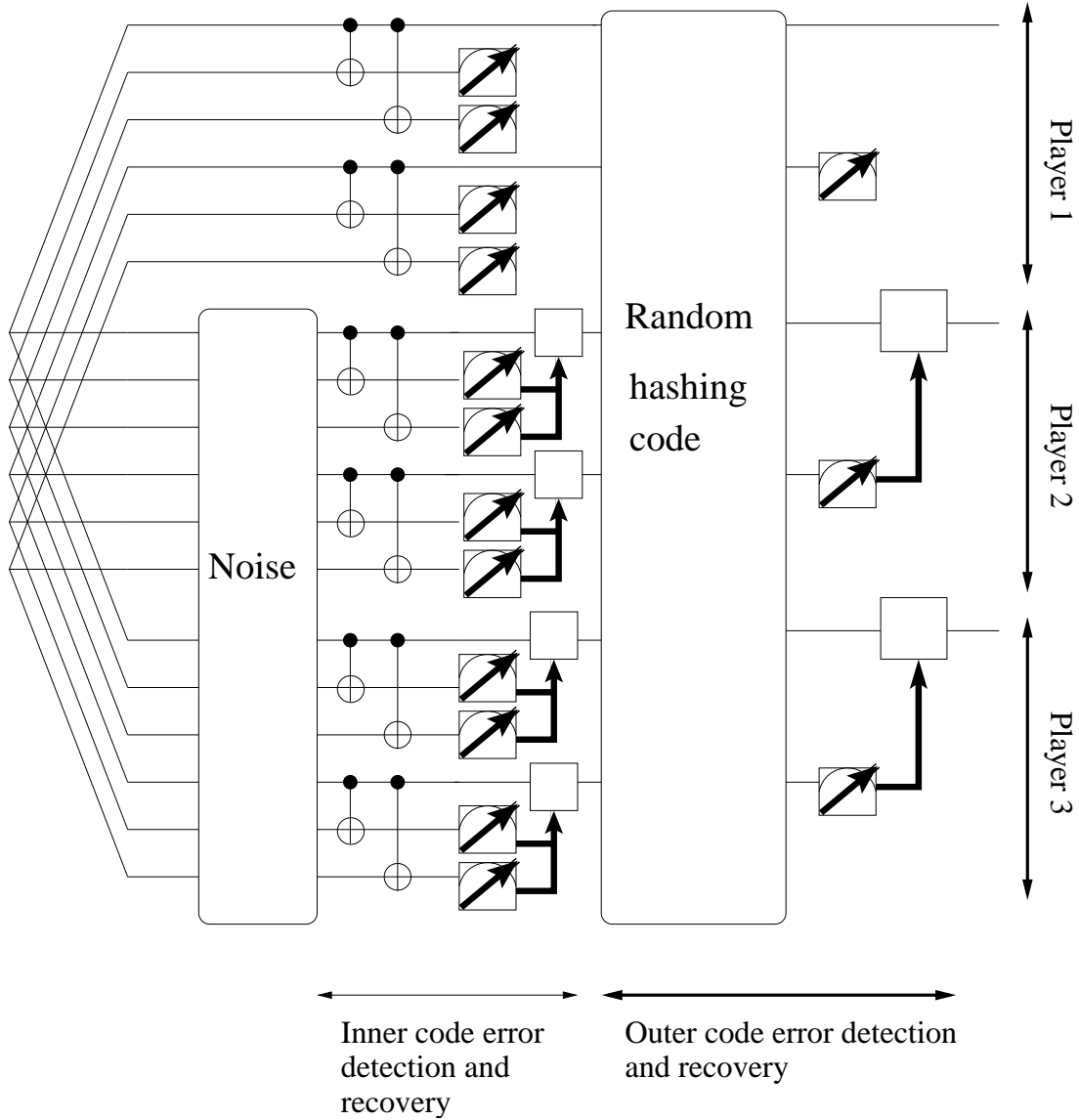}
\caption{Our three GHZ state distillation protocols for $m = n = 3$.  They all
 use the $[n,1,n]$ repetition code as their inner codes; and they differ by the
 kind of random hashing outer code used.
 \label{fig:our_protocol}
}
\end{figure}

Our three protocols are natural extensions of the Shor-Smolin concatenation 
procedure to the case of purifying GHZ states.  They all use
the same degenerate quantum code as the inner code.
Specifically, suppose the $m$ players share 
$N n$ copies of Werner state with $N\gg 1$. As shown in
Fig.~\ref{fig:our_protocol}, to distill
perfect GHZ state, each player applies the (classical) $[n, 1, n]$ repetition
code, whose stabilizer generators are
\begin{equation} \label{eq:GHZ_ZZ}
Z_0 Z_1, Z_0 Z_2, \ldots, Z_0 Z_{n-1} ~,
\end{equation}
to his/her own $n$ qubits. That is to say, they randomly partition the $N n$
shared noisy GHZ states into $N$ sets,
each containing $n$ noisy GHZ states. In each set ${\mathfrak S}$, they
randomly assign one of the noisy GHZ state as the source (and call
it the $0$th copy of $\ket{\Phi^{m+}}$
in the set) and the remaining $(n-1)$ noisy GHZ states as the targets (and call
them the $j$th copy of $\ket{\Phi^{m+}}$ in the set for $j = 1, 2, \ldots, n
- 1$).
They apply the MXOR operation to copies of $\ket{\Phi^{m+}}$ in each set
and then measure all the target GHZ states in the standard computational basis
while leaving all the source GHZ states un-measured. We
denote the syndrome and the remaining state in each set by
$\vec{\boldsymbol{s}}_{\mathfrak S}
\in GF(2)^{(n - 1) (m - 1)}$ and $(\delta_{\mathfrak S},
\boldsymbol{\gamma}_{\mathfrak S}) \in GF(2)^m$ respectively.
(Since the partition into $N$ sets is arbitrarily chosen and our subsequent
analysis only makes use of the statistical properties of the states in each
set, we drop the set label ${\mathfrak S}$ in all quantities to be analyzed
from now on.)

Our three protocols differ in the use of outer codes.  For the
first protocol, each player applies a (classical) random hashing code $[N, k_2, d_2]$ that corrects GHZ diagonal basis errors
to the $N$ remaining states (each coming from a different set ${\mathfrak S}$)
and exchanges the measurement results.
Clearly, this protocol is reduced to the Shor-Smolin concatenation
procedure~\cite{SS96} when $m = 2$.

For the second protocol, the players follow Maneva and Smolin's idea~\cite{MS00}
by using two (classical) random hashing codes, one to correct spin flip error and the other
to correct phase shift.  In this sense, the outer code used in our second
protocol is a random asymmetric Calderbank-Shor-Steane (CSS) code.  Whereas
players in our third protocol use the Chen and Lo's
generalization~\cite{CL07} as their outer code.  That is, the outer code is a
random asymmetric CSS code whose decoding circuit is carefully designed to
exploit the correlation between the bit string $b_i$.

In all the above three protocols, the players have to apply
the corresponding unitary transformation for the outer code to obtain the
purified GHZ states.  (See Fig.~\ref{fig:our_protocol}.)

Clearly, the yield of the first protocol is the Shor-Smolin capacity given by
Eqs.~(\ref{eq:D_SS}) and~(\ref{eq:Sx}).  And by applying Eq.~(\ref{eq:D1}) to
the noisy GHZ state to be fed into the outer code of the second protocol, we
conclude that the yield of the second protocol equals
\begin{equation}
 D_\textrm{MS} = \frac{1}{n} \left( 1 - \sum_{\vec{\boldsymbol{s}} \in
 GF(2)^{(m-1)(n-1)}} \textrm{Pr}(\vec{\boldsymbol{s}}) \left\{ \max_{1\leq i
 \leq m-1} \left[ H(\gamma_i | \vec{\boldsymbol{s}} ) \right] + H(\delta |
 \vec{\boldsymbol{s}}) \right\} \right) \label{eq:D_MS}
\end{equation}
where
\begin{equation}
 H ( \gamma_i | \vec{\boldsymbol{s}} ) = h ( \{ \sum_{\delta, \gamma_1,
 \ldots , \gamma_{i-1},\gamma_{i+1}, \ldots ,\gamma_{m-1}\in GF(2)}
 \textrm{Pr} ((\delta, \boldsymbol{\gamma}) |
 \vec{\boldsymbol{s}}) : \gamma_i \in GF(2) \} ) \label{eq:D_hDef1}
\end{equation}
for $i = 1,2,\ldots ,m-1$, and
\begin{equation}
 H ( \delta | \vec{\boldsymbol{s}} ) = h ( \{ \sum_{\boldsymbol{\gamma}\in
 GF(2)^{m-1}} \textrm{Pr} ((\delta, \boldsymbol{\gamma}) | \vec{\boldsymbol{s}})
 : \delta \in GF(2) \} ) . \label{eq:D_hDef2}
\end{equation}
Similarly, from Eq.~(\ref{eq:D2}), the yield of the third protocol is
\begin{equation}
 D_\textrm{CL} = \frac{1}{n} \left( 1 - \sum_{\vec{\boldsymbol{s}} \in
 GF(2)^{(m-1)(n-1)}} \textrm{Pr}(\vec{\boldsymbol{s}}) \left\{ \max_{1\leq i
 \leq m-1} \left[ H(\gamma_i | \gamma_{i-1}, \gamma_{i-2}, \ldots, \gamma_1,
 \vec{\boldsymbol{s}} ) \right]
 + H(\delta | \vec{\boldsymbol{s}})  - I (\delta;\boldsymbol{\gamma} |
 \vec{\boldsymbol{s}}) \right\} \right) \label{eq:D_CL}
\end{equation}
where
\begin{equation}
 H (\gamma_i | \gamma_{i-1},\gamma_{i-1},\ldots , \gamma_1,
 \vec{\boldsymbol{s}}) = h ( \{ \sum_{\delta,\gamma_{i+1},\gamma_{i+2},
 \ldots , \gamma_{m-1}\in GF(2)} \textrm{Pr} ((\delta,\boldsymbol{\gamma}) |
 \gamma_{i-1}, \gamma_{i-2} , \ldots , \gamma_1, \vec{\boldsymbol{s}}) :
 \gamma_i \in GF(2) \} ) \label{eq:D_hDef3}
\end{equation}
for $i=1,2,\ldots ,m-1$ and $I (\delta;\boldsymbol{\gamma} |
\vec{\boldsymbol{s}})$ is the conditional mutual information between
$\delta$, and $\boldsymbol{\gamma}$ given $\vec{\boldsymbol{s}}$.

\subsection{Evaluating the yields for Werner states for our protocols}
\label{subsec:evaluation}
To analyze the performance of our three protocols when applied to Werner states,
we first have to calculate the distribution of outcomes after passing the
Werner states through the inner repetition code.
For an arbitrary but fixed set
${\mathfrak S}$, using
the compact notation introduced in Sec.\ref{sec:GHZ_re}, we denote the error
experienced by the $j$th copy of $\ket{\Phi^{m+}}$ in this set by
$(\beta_j, \boldsymbol{\alpha}_j)$ for $j = 0, 1, \ldots , n - 1$. 
After decoding the inner code, namely, the $[n, 1, n]$ repetition
code whose generators of the stabilizer are written down in
Eq.~(\ref{eq:GHZ_ZZ}), the syndrome $\vec{\boldsymbol{s}} \equiv
(\boldsymbol{s}_1,\boldsymbol{s}_2, \ldots \boldsymbol{s}_{n-1}) \in
GF(2)^{(m-1)(n-1)}$ obtained obeys
\begin{equation}
\boldsymbol{s}_j \equiv \boldsymbol{\alpha}_j - \boldsymbol{\alpha}_0 \in
GF(2)^{m-1} \label{eq:alpha_j_relation}
\end{equation}
for all $1\leq j\leq n-1$.  Furthermore, the remaining state shared among
the $m$ players is
\begin{equation} 
(\delta, \boldsymbol{\gamma}) = 
(\sum_{j = 0}^{n - 1} \beta_j, \boldsymbol{\alpha}_0) ~.
\end{equation}
To simplify notation in our subsequent discussions, we define
\begin{equation}
 \boldsymbol{s}_0 = \boldsymbol{0}
\end{equation}
so that Eq.~(\ref{eq:alpha_j_relation}) is also valid for $j = 0$.

To evaluate the capacity for each of our three protocols, we first have to
calculate the
conditional probabilities $\text{Pr}( (\delta, \boldsymbol{\gamma}) |
\vec{\boldsymbol{s}})$ and
$\textrm{Pr}( (\delta,\boldsymbol{\gamma}) | \gamma_1,\gamma_2, \ldots,
\gamma_{i-1},\vec{\boldsymbol{s}})$ in Eqs.~(\ref{eq:Sx})
and~(\ref{eq:D_CL}), respectively.
We begin by computing the probability
$\text{Pr}((\delta,\boldsymbol{\gamma}) \wedge
\vec{\boldsymbol{s}})$ that the source state has experienced the error
$(\delta,\boldsymbol{\gamma}) = (\sum_{j=0}^{n-1}
\beta_j,\boldsymbol{\alpha}_0)$
after the decoding transformation of the repetition code in
Eq.~(\ref{eq:GHZ_ZZ}) and that the error syndrome for the repetition
code is $\vec{\boldsymbol{s}} \in GF(2)^{(m - 1) (n - 1) }$.  Clearly,
\begin{equation}
\text{Pr}((\delta,\boldsymbol{\gamma}) \wedge \vec{\boldsymbol{s}}) =
\text{Pr}({\mathfrak E}(\vec{\boldsymbol{s}},\delta,\boldsymbol{\gamma}))
\label{eq:pr_a}
\end{equation}
where
\begin{equation}
{\mathfrak E}(\vec{\boldsymbol{s}}, \delta, \boldsymbol{\gamma}) \equiv
\{ ( \beta_j,  \boldsymbol{\alpha}_j)_{j=0}^{n-1} \in GF(2)^{m n} : 
 \delta = \sum_{j=0}^{n - 1} \beta_j , \boldsymbol{\gamma} =
 \boldsymbol{\alpha}_0, \boldsymbol{\alpha}_\ell =  \boldsymbol{s}_\ell +
 \boldsymbol{\alpha}_0 \enspace \text{for} \enspace \ell = 1, 2, \ldots, n -1\}
 ~. \label{eq:E_set_def}
\end{equation}
Since the repetition code and our decoding transformation are highly
symmetric in the sense that they are invariant under relabeling of qubits,
it is not surprising that the set
${\mathfrak E}(\vec{\boldsymbol{s}},\delta,\boldsymbol{\gamma})$ is invariant
under permutation of phase errors.  That is to say,
$(\beta_j,\boldsymbol{\alpha}_j)_{j=0}^{n-1} \in
{\mathfrak E}(\vec{\boldsymbol{s}},\delta,\boldsymbol{\gamma})$ if and only if
$(\beta_{\pi(j)},\boldsymbol{\alpha}_j)_{j=0}^{n-1} \in
{\mathfrak E}(\vec{\boldsymbol{s}},\delta,\boldsymbol{\gamma})$ where $\pi$ is
a permutation of $\{ 0,1,\ldots ,n-1 \}$.

\subsubsection{Finding $\text{Pr}((\delta,\boldsymbol{\gamma}) \wedge
\vec{\boldsymbol{s}})$}
We proceed by introducing the concepts of depolarization weight and
depolarization weight enumerator similar to the
ones proposed by DiVincenzo \emph{et al.}~\cite{DSS98}.
Let $(\beta_j, \boldsymbol{\alpha}_j)$ be the
state of the $j$th noisy GHZ state shared among the $m$ players.  The
{\bf depolarization weight} of the order $n$-tuple
$(\beta_j,\boldsymbol{\alpha}_j)_{j=0}^{n-1} \in GF(2)^{m n}$ is
defined as its Hamming weight by regarding this $n$-tuple as a vector of
elements in $GF(2)^m$.  In other words,
\begin{equation}
\text{wt}\left((\beta_j,\boldsymbol{\alpha}_j)_{j=0}^{n-1}
\right) = | \{ j \in \{ 0,1,
\ldots , n-1 \} : (\beta_j, \boldsymbol{\alpha}_j) \neq
(0, \boldsymbol{0}) \}| ~.
\end{equation}
Physically, the depolarization weight measures the number of shared GHZ states
that experienced an error; thus, it is invariant under permutation of the $n$
possibly imperfect GHZ states.
Since a GHZ state has equal probability of having each type of error after
passing through a depolarizing channel, there is an equal probability
for the $n$ depolarized GHZ states to experience errors with the same
depolarization weight.
Thus, we may find the probability $\text{Pr}((\delta,\boldsymbol{\gamma})
\wedge \vec{\boldsymbol{s}})$ by studying the
{\bf depolarization weight enumerator}
$w({\mathfrak E}(\vec{\boldsymbol{s}},\delta,\boldsymbol{\gamma});x,y)$ where
\begin{equation}
w({\mathfrak A};x,y) = \sum_{\vec{\boldsymbol{a}}\in {\mathfrak A}}
x^{\text{wt}(\vec{\boldsymbol{a}})} y^{n - \text{wt}(\vec{\boldsymbol{a}})} ~.
\end{equation}
The depolarization weight enumerator of a set is a natural generalization of
the concept of weight enumerator of a code.

Finding an explicit expression for the above depolarization weight enumerator
for an arbitrary set or coset is a very difficult task.  It is the high degree
of symmetry in the repetition code that makes this task possible.  In fact, one
may transform a state in
${\mathfrak E}(\vec{\boldsymbol{s}},\delta,\boldsymbol{\gamma})$ to another
state in the same set by applying phase shifts to a few qubits.

By counting the number of different possible combinations of $(\beta_j,
\boldsymbol{\alpha}_j)$'s subjected to the constraint that
$(\beta_j,\boldsymbol{\alpha}_j)_{j=0}^{n-1} \in {\mathfrak E}
(\vec{\boldsymbol{s}},\delta,\boldsymbol{\gamma})$, we have
\begin{equation} \label{eq:wx1}
w({\mathfrak E}(\vec{\boldsymbol{s}},\delta,\boldsymbol{\gamma});x,y) =
\mathop{{\sum}'} \frac{n!}{ \displaystyle
\prod_{\genfrac{}{}{0pt}{}{i \in GF(2),}{\boldsymbol{t} \in GF(2)^{m-1}} }
a_{i, \boldsymbol{t}}!} \enspace x^{n - a_{0, \boldsymbol{0}}} y^{a_{0,
\boldsymbol{0}}}
\end{equation}
where the primed sum is over all $a_{i,\boldsymbol{t}}$'s satisfying the
constraints
\begin{equation}
a_{i, \boldsymbol{t}} \geq 0 \enspace \forall i, \boldsymbol{t} ~,
\label{eq:con_a}
\end{equation}
\begin{equation}
\sum_{\genfrac{}{}{0pt}{}{i \in GF(2),}{\boldsymbol{t} \in GF(2)^{m-1}}}
a_{i, \boldsymbol{t}} = n ~,
\label{eq:con_b}
\end{equation}
\begin{equation}
\sum_{i\in GF(2)} a_{i,\boldsymbol{t}} = | \{ j \in \{ 0,1,\ldots,
n-1 \} : \boldsymbol{s}_j + \boldsymbol{\gamma} = \boldsymbol{t} \} |
\enspace \forall \boldsymbol{t}
\label{eq:con_c}
\end{equation}
and
\begin{equation}\label{eq:con1}
\sum_{\genfrac{}{}{0pt}{}{i \in GF(2),}{\boldsymbol{t} \in GF(2)^{m-1}}} i
\, a_{i,\boldsymbol{t}} = \delta ~.
\end{equation}
Note that the symbol $a_{i,\boldsymbol{t}}$ in the above equations can be
interpreted as the number of GHZ states that experienced the error
$(i,\boldsymbol{t})$ before the commencement of our distillation protocol.

Let
\begin{equation}
k \equiv k(\vec{\boldsymbol{s}},\boldsymbol{\gamma}) =
|\{ j \in \{ 0,1,\ldots , n-1 \} : \boldsymbol{s}_j + \boldsymbol{\gamma} \neq
\boldsymbol{0} \} | \label{eq:k_def}
\end{equation}
be the number of qubits having spin flip for each element in
${\mathfrak E}(\vec{\boldsymbol{s}},\delta,\boldsymbol{\gamma})$.
We have two cases to consider.
\par\noindent
Case~(a) $k > 0$: That is, there exists $\ell$ such that
$\boldsymbol{s}_\ell + \boldsymbol{\gamma} \neq \boldsymbol{0}$.  Hence,
$\text{wt}((\beta_j, \boldsymbol{s}_j + \boldsymbol{\gamma})_{j=0}^{n-1})$ is
independent of the value of $\beta_\ell
\in GF(2)$.  In addition, by regarding the equation $\sum_{j=0}^{n-1} \beta_j =
\delta$ as a bijection relating $\beta_\ell \in GF(2)$ and
$\delta\in GF(2)$, we conclude that the depolarization weight enumerator
$w({\mathfrak E}(\vec{\boldsymbol{s}},\delta,\boldsymbol{\gamma});x,y)$ is
independent of the value of $\delta\in GF(2)$.  Hence,
\begin{equation}
w({\mathfrak E}(\vec{\boldsymbol{s}},\delta,\boldsymbol{\gamma});x,y) =
\frac{1}{2} \mathop{{\sum}''}
\frac{n!}{\displaystyle \prod_{\genfrac{}{}{0pt}{}{i \in GF(2),}{\boldsymbol{t}
\in GF(2)^{m-1}}}  a_{i, \boldsymbol{t}}!} \enspace
x^{n - a_{0, \boldsymbol{0}}} y^{a_{0,\boldsymbol{0}}}
\end{equation}
where the double primed sum is over all $a_{i,\boldsymbol{t}}$'s satisfying
constraints Eq.~(\ref{eq:con_a})--(\ref{eq:con_c}) only. Consequently,
\begin{eqnarray}
w({\mathfrak E}(\vec{\boldsymbol{s}},\delta,\boldsymbol{\gamma});x,y) &=& 
\frac{1}{2} \sum_{ \{ a_{i, \boldsymbol{t}} \} }
\frac{k!}{\displaystyle \prod_{\genfrac{}{}{0pt}{}{i\in GF(2),}{\boldsymbol{t}
\in {GF(2)^{m-1}}\setminus \{ \boldsymbol{0} \}}} a_{i,\boldsymbol{t}}}
\enspace {n - k \choose a_{0, \boldsymbol{0}}}
\enspace x^{n -a_{0, \boldsymbol{0}}} y^{a_{0,\boldsymbol{0}}}
\nonumber \\
&=& 2^{k -1} \sum_{a_{0, \boldsymbol{0}}}
{n - k \choose a_{0, \boldsymbol{0}}} \enspace
x^{n - a_{0, \boldsymbol{0}}} y^{a_{0,\boldsymbol{0}}} \nonumber \\
&=& 2^{k-1} x^k (x+y)^{n-k} ~.
\label{eq:dweight1}
\end{eqnarray}

\par\noindent
Case~(b) $k = 0$: That is, $\boldsymbol{s}_j + \boldsymbol{\gamma} =
\boldsymbol{0}$ for all $j$
so that phase shift is the only type of error a GHZ state may experience.
In this case, the union of disjoint sets
$\bigcup_{\delta\in GF(2)} {\mathfrak E}(\vec{\boldsymbol{s}},\delta,
\boldsymbol{\gamma})$ is equal to the set of all possible phase errors
experienced by the $n$ shared GHZ states. As a result,
\begin{equation}
\sum_{\delta\in GF(2)} w({\mathfrak E}(
\vec{\boldsymbol{s}},\delta,\boldsymbol{\gamma});x,y) =
w(\bigcup_{\delta\in GF(2)} {\mathfrak E}(\vec{\boldsymbol{s}},\delta,
\boldsymbol{\gamma});x,y) = \sum_i {n \choose i}
\enspace x^i y^{n-i} = (x+y)^n ~.
\label{eq:dweight2}
\end{equation}
Similarly,
\begin{equation}
\sum_{\delta\in GF(2)} (-1)^\delta w({\mathfrak E}(
\vec{\boldsymbol{s}},\delta,\boldsymbol{\gamma});x,y) =
w(\bigcup_{\delta\in GF(2)} {\mathfrak E}(\vec{\boldsymbol{s}},\delta,
\boldsymbol{\gamma});-x,y) = \sum_i {n \choose i} \enspace (-x)^i y^{n-i} =
(y-x)^n ~.
\label{eq:dweight3}
\end{equation}
(Note that the validity of the above equation follows from the observation
that $\delta = 1$
if and only if the number of qubits having phase shift error before the
commencement of our protocol is odd.
Moreover, the coefficient of $w(\{ \vec{\boldsymbol{a}} \};-x,y)$ is negative if
and only if $wt(\vec{\boldsymbol{a}})$ is odd.)
From Eqs.~(\ref{eq:dweight1})--(\ref{eq:dweight3}), we conclude that
\begin{equation}
w({\mathfrak E}(\vec{\boldsymbol{s}},\delta,\boldsymbol{\gamma});x,y) \equiv
w(k,\delta;x,y) = \left\{ 
\begin{array}{ll}
 \displaystyle
 2^{k - 1} x^k (x+y)^{n-k} & \enspace\text{if} \enspace 0 < k \leq n, \\
 \\
 \displaystyle
 \frac{1}{2} \left[ (x + y)^n + (y - x)^n \right] & \enspace\text{if}
  \enspace k = 0 \enspace \text{and} \enspace \delta = 0, \\
 \\
 \displaystyle
 \frac{1}{2} \left[ (x + y)^n - (y - x)^n \right] & \enspace\text{if}
  \enspace k = 0 \enspace \text{and} \enspace \delta \neq 0,
\end{array}
\right. \label{eq:w_k_delta_x}
\end{equation}
where $k = k(\vec{\boldsymbol{s}},\boldsymbol{\gamma})$ is the number of GHZ
states that experienced some kind of spin flip for each of the state in
${\mathfrak E}(\vec{\boldsymbol{s}},\delta,\boldsymbol{\gamma})$ as defined by
Eq.~(\ref{eq:k_def}).

Recall that ${\mathfrak E}( \vec{\boldsymbol{s}},\delta,\boldsymbol{\gamma})$ is
invariant under permutation of phase errors among the $n$ GHZ states.  Moreover,
both the depolarization weight and the value of $k(\vec{\boldsymbol{s}},
\boldsymbol{\gamma})$ are invariant under permutation of the $n$ GHZ states.
So, it is not surprising that the depolarizing weight enumerator
$w({\mathfrak E} (\vec{\boldsymbol{s}},\delta,\boldsymbol{\gamma});x,y)$
depends only on the values of $k$ and $\delta$.  Therefore, our
shorthand notation $w(k,\delta;x)$ makes sense.

From Eq.~(\ref{eq:pr_a}) and by substituting
$x = (1-F)/(2^m - 1)$, $y = F$ into Eq.~(\ref{eq:w_k_delta_x}), we find that
\begin{equation}
 \text{Pr}((\delta,\boldsymbol{\gamma}) \wedge \vec{\boldsymbol{s}}) = \left\{
 \begin{array}{ll}
  \displaystyle \frac{2^{k-1} (1-F)^k (2^m F - 2F + 1)^{n-k}}{(2^m - 1)^n}
  & \enspace \text{if} \enspace 0 < k \leq n , \\
  \\
  \displaystyle \frac{(2^m F - 2F + 1)^n + (2^m F - 1)^n}{2(2^m - 1)^n} &
   \enspace \text{if} \enspace k = 0 \enspace \text{and} \enspace \delta = 0 ,
  \\
  \\
  \displaystyle \frac{(2^m F - 2F + 1)^n - (2^m F - 1)^n}{2(2^m - 1)^n} &
   \enspace \text{if} \enspace k = 0 \enspace \text{and} \enspace \delta \neq
   0 .
 \end{array} \right.
 \label{eq:pr_k_delta}
\end{equation}
for a depolarizing channel with fidelity $F$.
Note that by fixing the number of players $m$, the number of noisy GHZ states
shared between the players $n$ and the fidelity of the depolarizing channel
$F$, the probability $\text{Pr}((\delta,\boldsymbol{\gamma}) \wedge
\vec{\boldsymbol{s}})$ can take on at most $(n+2)$ different values.

\subsubsection{Finding $\text{Pr}((\delta,\boldsymbol{\gamma})|
\vec{\boldsymbol{s}})$}
\label{subsubsec:basic_probability}
Clearly
\begin{equation}
 \text{Pr} (\vec{\boldsymbol{s}}) = \sum_{\boldsymbol{t}\in GF(2)^{m-1}}
 \text{Pr} (\vec{\boldsymbol{s}} \wedge \boldsymbol{t})
\end{equation}
where $\text{Pr}(\vec{\boldsymbol{s}} \wedge \boldsymbol{t})$ is the
probability that the error experienced by the $n$ noisy GHZ states is
$(\beta_j,\boldsymbol{s}_j +
\boldsymbol{t})_{j=0}^{n-1}$ for $\beta_j \in GF(2)$.  For a
depolarizing channel with fidelity $F$,
\begin{eqnarray}
 \text{Pr} (\vec{\boldsymbol{s}}) & = &
 \sum_{\boldsymbol{t}\in GF(2)^{m-1}} \left[ \frac{2(1-F)}{2^m - 1}
 \right]^{k(\vec{\boldsymbol{s}},\boldsymbol{t})} \enspace \left( F +
 \frac{1-F}{2^m - 1} \right)^{n - k(\vec{\boldsymbol{s}},\boldsymbol{t})}
 \nonumber \\
 & = & \frac{1}{(2^m - 1)^n} \sum_{i=0}^n f_{\vec{\boldsymbol{s}}} (i) 2^i
 (1-F)^i (2^m F - 2F + 1)^{n-i}
 \label{eq:pr_s_sum}
\end{eqnarray}
where
\begin{equation}
f_{\vec{\boldsymbol{s}}} (i) = |\{ \boldsymbol{t} \in GF(2)^{m - 1} :
k(\vec{\boldsymbol{s}},\boldsymbol{t}) = i \}| ~.
\end{equation}
Therefore,
\begin{equation}
 \text{Pr}((\delta,\boldsymbol{\gamma}) | \vec{\boldsymbol{s}}) = \left\{
 \begin{array}{ll}
  \displaystyle \frac{2^{k-1}(1-F)^k (2^m F - 2F + 1)^{n-k}}{\sum_i
   f_{\vec{\boldsymbol{s}}} (i) 2^i (1-F)^i (2^m F - 2F + 1)^{n-i}} &
   \enspace \text{if} \enspace 0 < k \leq n , \\
  \\
  \displaystyle \frac{(2^m F - 2F + 1)^n + (2^m F - 1)^n}{2 \sum_i
   f_{\vec{\boldsymbol{s}}} (i) 2^i (1-F)^i (2^m F - 2F + 1)^{n-i}} &
   \enspace \text{if} \enspace k = 0 \enspace \text{and} \enspace \delta =
   0 , \\
  \\
  \displaystyle \frac{(2^m F - 2F + 1)^n - (2^m F - 1)^n}{2 \sum_i
   f_{\vec{\boldsymbol{s}}} (i) 2^i (1-F)^i (2^m F - 2F + 1)^{n-i}} &
   \enspace \text{if} \enspace k = 0 \enspace \text{and} \enspace \delta \neq
   0 .
 \end{array} \right.
 \label{eq:p_delta_gamma_cond}
\end{equation}
So combined with Eqs.~(\ref{eq:D_SS}) and~(\ref{eq:Sx}), we have a working
expression for $h(\{\text{Pr}((\delta,\boldsymbol{\gamma})|
\vec{\boldsymbol{s}}) :
(\delta, \boldsymbol{\gamma}) \in GF(2)^m \})$ and hence $D_\textrm{SS}$.
While combined with Eqs.~(\ref{eq:D_MS})--(\ref{eq:D_hDef2}), we have a working
expression for $h(\gamma_i | \vec{\boldsymbol{s}})$, $h(\delta |
\vec{\boldsymbol{s}})$ and hence $D_\textrm{MS}$.
In the calculation of $h(\gamma_i | \gamma_{i-1}, \gamma_{i-2}, \ldots ,
\gamma_1 , \vec{\boldsymbol{s}})$, we need to first compute the probability
$\textrm{Pr}( (\delta, \boldsymbol{\gamma}) | \gamma_{i-1},\gamma_{i-2},
\ldots , \gamma_1, \vec{\boldsymbol{s}})$.
As for repetition code, Eq.~(\ref{eq:alpha_j_relation}) tells us that
the kind of spin flip error experienced by the $j$th GHZ state
$\boldsymbol{\alpha}_j$ is known once $\boldsymbol{\gamma}$ and
$\vec{\boldsymbol{s}}$ are fixed.  Hence,
$\textrm{Pr}( (\delta, \boldsymbol{\gamma}) | \gamma_{i-1},\gamma_{i-2},
\ldots , \gamma_1, \vec{\boldsymbol{s}})$ is also given by
Eq.~(\ref{eq:p_delta_gamma_cond}).  Consequently, using
Eqs.~(\ref{eq:D_hDef2})--(\ref{eq:D_hDef3}), we get a working
expression for $h(\gamma_i | \gamma_{i-1},\ldots ,\gamma_1,
\vec{\boldsymbol{s}})$, $h(\delta | \vec{\boldsymbol{s}})$, $I(\delta;
\boldsymbol{\gamma} | \vec{\boldsymbol{s}})$ and hence $D_\textrm{CL}$.

\subsubsection{Complexity issue on the computation of $D_\textrm{SS}$,
 $D_\textrm{MS}$ and $D_\textrm{CL}$}
\label{subsubsec:complexity}

Apparently computing $D_\textrm{SS}$, $D_\textrm{MS}$ and $D_\textrm{CL}$ using
Eqs.~(\ref{eq:D_SS}), (\ref{eq:Sx}), (\ref{eq:D_MS})--(\ref{eq:D_hDef3}),
(\ref{eq:pr_s_sum})--(\ref{eq:p_delta_gamma_cond})
are extremely inefficient as the sum on $\vec{\boldsymbol{s}}$ may take on
$2^{m(n-1)}$ possible values.  Nonetheless, the numerical values of many
terms in the R.H.S. of Eq.~(\ref{eq:Sx}) are the same because the
$\vec{\boldsymbol{s}}$ dependence of $\text{Pr}
(\vec{\boldsymbol{s}})$ and $\text{Pr}((\delta,\boldsymbol{\gamma}) |
\vec{\boldsymbol{s}})$ come indirectly from the distribution of $\{
k(\vec{\boldsymbol{s}},\boldsymbol{t}) : \boldsymbol{t} \in GF(2)^{m-1} \}$.
Note that there are at most $\sum_{i=0}^{2^{m - 1}} {\mathcal P}_i (n)$
different possible distributions for $\{ k(\vec{\boldsymbol{s}},\boldsymbol{t})
: \boldsymbol{t} \in GF(2)^{m-1} \}$ where ${\mathcal P}_i (n)$ denotes the
number of ways to express $n$ as a sum of exactly $i$ positive integers.
Moreover, $\mathcal{P}_{i}(n)$ scales as $\exp(\pi \sqrt{2n/3}) / 4n\sqrt{3}$
in the large $n$ limit~\cite{GE76}.
Consequently, for a fixed $m$, we may regroup the sum Eq.~(\ref{eq:Sx}) so as
to compute $S_X$ by summing only sub-exponential in $n$ terms.
Although this is not a polynomial time in $n$ algorithm, it is good
enough to obtain the numerical values for $S_X$ and hence the yield of
our first protocol, namely, the Shor-Smolin capacity $D_\textrm{SS}$ for a
reasonably large number of $n$.
By the same token, the yields of our second and third protocols, namely
$D_\textrm{MS}$ and $D_\textrm{CL}$ respectively, can also be
computed in sub-exponential time in $n$.

\subsection{Performance of our three schemes}
\label{subsec:performance}

We study the performance of our three protocols by studying the yield
as a function of the channel fidelity $F$.  In particular, we
would like to find the threshold fidelity, namely, the minimum fidelity above
which $D > 0$, as
a function of the number of players $m$ and the repetition codeword size $n$.
And we denote the threshold fidelities for our first, second and third
protocols by
$F_\textrm{min}^\textrm{SS} (m,n)$, $F_\textrm{min}^\textrm{MS} (m,n)$ and
$F_\textrm{min}^\textrm{CL} (m,n)$, respectively.

\subsubsection{Subtlety in the computation of threshold fidelities}
\label{subsubsec:threshold_fidelities_computation}

Finding the values of
$F_\textrm{min}^\textrm{SS} (m,n)$, $F_\textrm{min}^\textrm{MS} (m,n)$ and
$F_\textrm{min}^\textrm{CL} (m,n)$ requires extra care.
Let us explain why for the case of $F_\textrm{min}^\textrm{SS} (m,n)$.  And
the reason for the other two cases are similar.

Since $S_X$ is a continuous function of the channel fidelity $F$,
Eq.~(\ref{eq:D_SS}) implies that
$F_\text{min}^\textrm{SS}(m,n)$ is the root of the equation $S_X = 1$. Note
that
\begin{eqnarray}
 1 - S_X & = & \text{Pr}(\vec{\boldsymbol{0}}) [ 1 -
  h(\{ \text{Pr}((\delta,\boldsymbol{\gamma})|\vec{\boldsymbol{0}}) :
  (\delta,\boldsymbol{\gamma}) \in GF(2)^m \}) ]
 \nonumber \\
 & & ~~ - \sum_{\vec{\boldsymbol{s}}\neq \vec{\boldsymbol{0}}} \text{Pr}
  (\vec{\boldsymbol{s}}) [
  h(\{ \text{Pr}((\delta,\boldsymbol{\gamma})|\vec{\boldsymbol{s}}) :
  (\delta,\boldsymbol{\gamma}) \in GF(2)^m \}) - 1 ] ~. \label{eq:Sx_alt}
\end{eqnarray}
From Eq.~(\ref{eq:p_delta_gamma_cond}), we know that for $F\gg 1/2$,
$h(\{ \text{Pr}((\delta,\boldsymbol{\gamma})|\vec{\boldsymbol{s}}) : (\delta,
\boldsymbol{\gamma}) \in GF(2)^m \})$ is less (greater) than $1$ if
$\vec{\boldsymbol{s}} = \vec{\boldsymbol{0}}$
($\vec{\boldsymbol{s}} \neq \vec{\boldsymbol{0}}$). More importantly,
for a fixed $m$, $\lim_{n\rightarrow\infty} h(\{ \text{Pr}((\delta,
\boldsymbol{\gamma})| \vec{\boldsymbol{s}}) : (\delta,\boldsymbol{\gamma}) \in
GF(2)^m \}) = 1^- (1^+)$ for $\vec{\boldsymbol{s}} = \vec{\boldsymbol{0}}$
($\vec{\boldsymbol{s}} \neq \vec{\boldsymbol{0}}$).
Thus, Eq.~(\ref{eq:Sx_alt}) shows that $1 - S_X$ is the difference
between two small positive terms.  This makes the computation of
$F_\text{min}^\textrm{SS}(m,n)$ together with the analysis of its trend as a
function of $m$ and $n$, particularly for a large $n$, difficult.
Even worse, for $F<1$ and for a
sufficiently large $n$, the errors experienced by the noisy GHZ states
$(\delta,\boldsymbol{\gamma})_{i=0}^{n-1}$ satisfying $\vec{\boldsymbol{s}} =
\vec{\boldsymbol{0}}$ are not in the typical set.  Actually, we found that for
$F$ close to $F_\text{min}^\textrm{SS}(m,n)$, the dominant terms in the R.H.S.
of Eq.~(\ref{eq:Sx_alt}) almost always correspond to atypical errors
experienced by the GHZ states.

In spite of these difficulties, we are able to accurately compute the yield of
our first protocol $D_\textrm{SS}$, namely, the Shor-Smolin capacity, as a
function of the channel fidelity $F$ for the classical
$[n,1,n]$ repetition code acting on the $\ket{\Phi^{m+}}$'s.  And from this,
we can deduce the correct threshold fidelity for our first protocol
$F_\textrm{min}^\textrm{SS} (m,n)$ as a function of the number of players $m$
and the number of shared noisy GHZ states $n$.  The trick is to use
rational number arithmetic to obtain an expression for $S_X$ for a given
rational number $F$ before converting this expression to an approximate
real number.  The same trick also enables us to obtain accurate values for
$F_\textrm{min}^\textrm{MS}(m,n)$ and $F_\textrm{min}^\textrm{CL}(m,n)$,
namely, the threshold fidelities of our second and third protocols.

\subsubsection{The superior performances of our three protocols}
\label{subsubsec:performances}

\begin{figure}[t]
\centering\includegraphics[width=14.5cm]{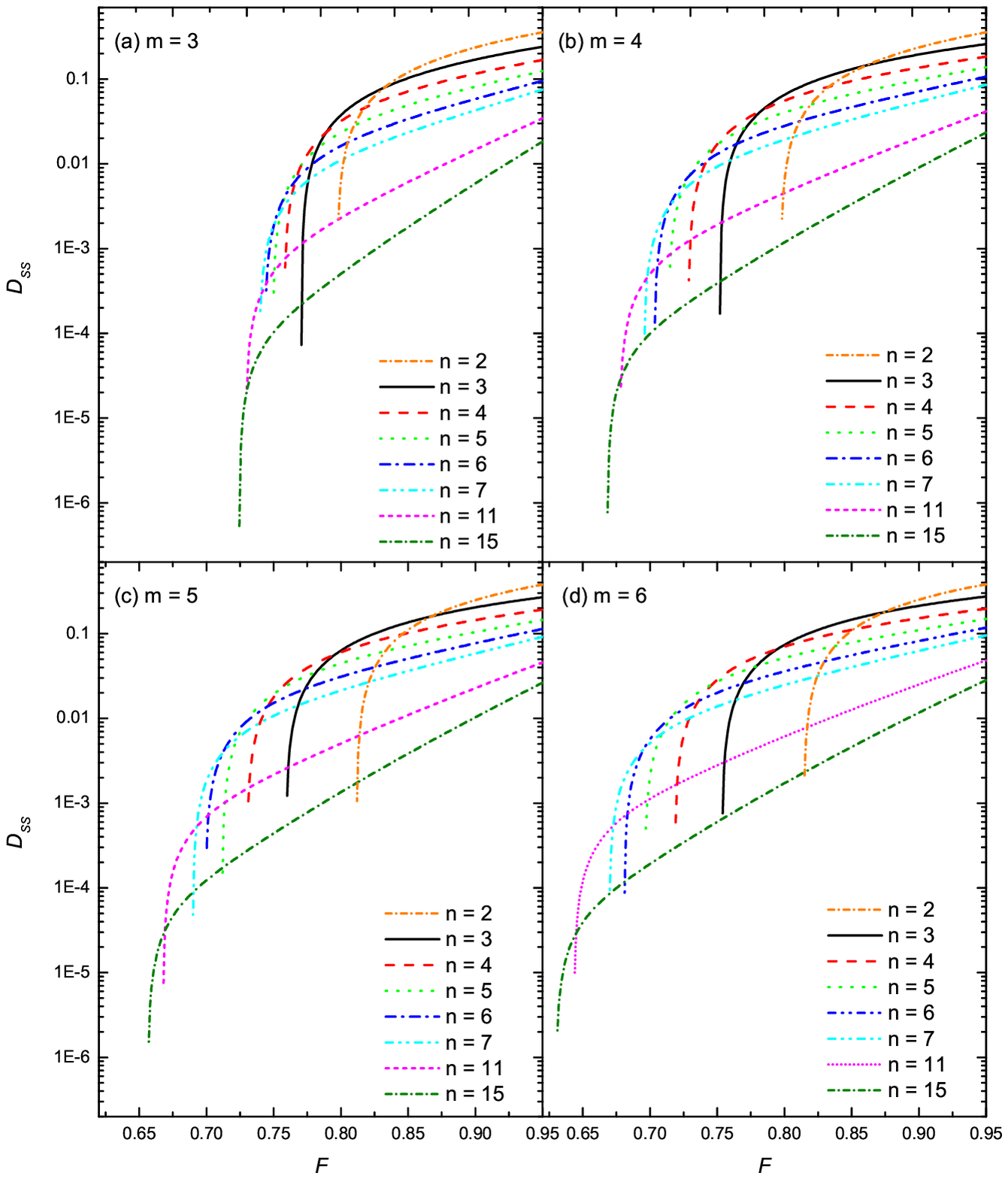}
 \caption{The yield $D_\textrm{SS}$ of our first protocol for distilling
  $\ket{\Phi^{m+}}$ after passing through a depolarizing channel of fidelity
  $F$ using the classical repetition code $[n, 1, n]$ as the inner code for
  various $n$ when (a)~$m = 2$, (b)~$m = 3$, (c)~$m = 4$ and (d)~$m = 5$.}
\label{fig:SS_m2-m5}
\end{figure}

\begin{figure}[t]
\centering\includegraphics[width=14.5cm]{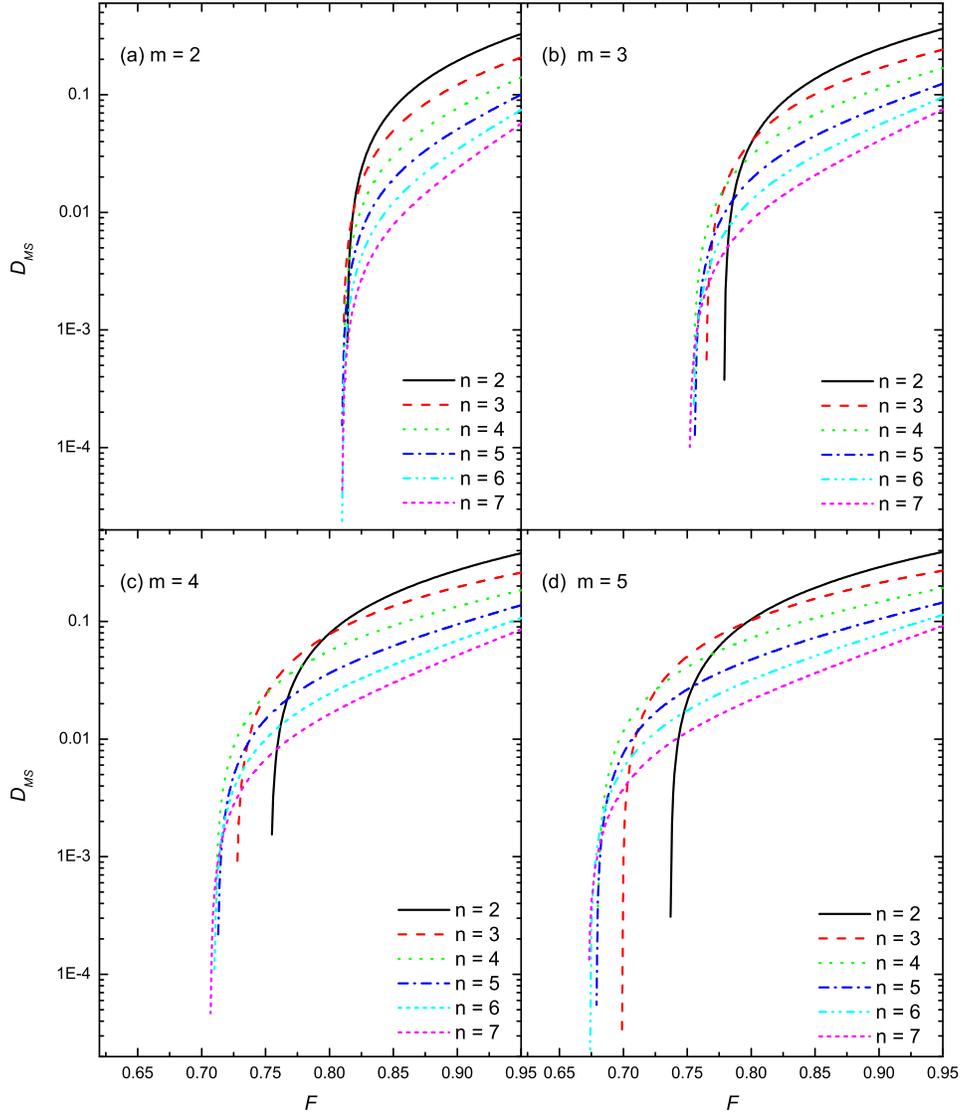}
 \caption{The yield $D_\textrm{MS}$ of our second protocol using the same sets
  of parameters as in Fig.~\ref{fig:SS_m2-m5}.}
\label{fig:MS_m2-m5}
\end{figure}

\begin{figure}[t]
\centering\includegraphics[width=14.5cm]{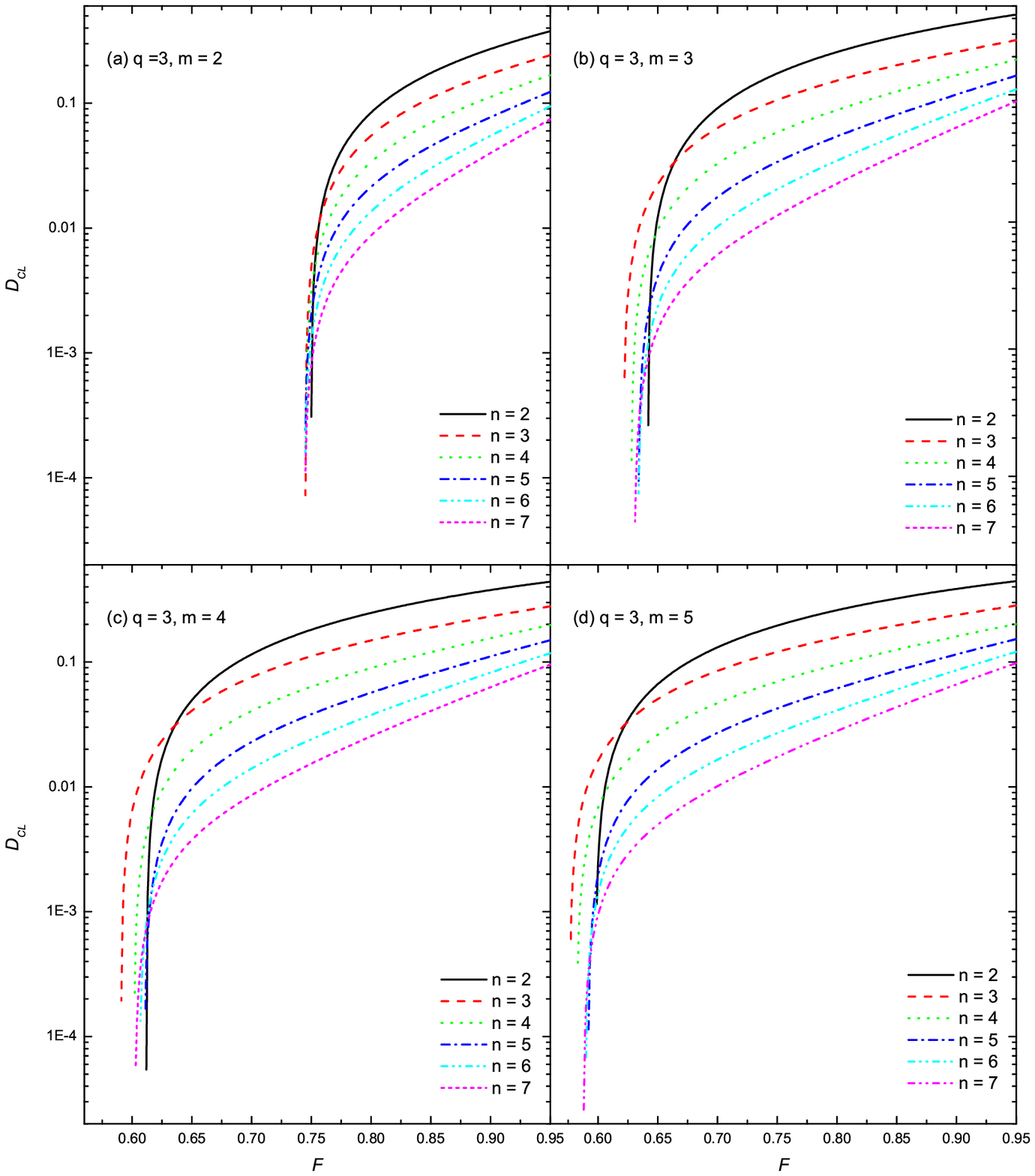}
 \caption{The yield $D_\textrm{CL}$ of our third protocol using the same sets
  of parameters as in Fig.~\ref{fig:SS_m2-m5}.}
\label{fig:CL_m2-m5}
\end{figure}

The yields of our three protocols are shown in
Figs.~\ref{fig:SS_m2-m5}--\ref{fig:CL_m2-m5}; and the corresponding
threshold fidelities are tabulated in
Tables~\ref{tb:SS_m2-m6}--\ref{tb:CL_m2-m6}.
By comparing these tables with the second column of Table~\ref{tb:pw},
we make the most important conclusion of this paper: for the
multipartite case ($m \geq 3$) and for any number of shared GHZ states $n$,
the error-tolerant capability of our third protocol is strictly better than our
second, which is in turn strictly better than that of our first.
And under the same conditions, the error-tolerant capability of our first
protocol is already better than the best scheme in literature before this work.
So once again, we show the powerfulness and usefulness of degenerate codes in
one-way entanglement distillation.

\begin{table}[t]
\begin{tabular}{||c|c|ccccccccc||}
\hline\hline
\multicolumn{2}{||c|}{} & \multicolumn{9}{c||}{$m$} \\
\cline{3-11}
\multicolumn{2}{||c|}{$F_\textrm{min}^\textrm{SS}(m, n)$} &
2 & ~~~~ & 3 & ~~~~ & 4 & ~~~~ & 5 & ~~~~ & 6 \\
\hline
& 2 & 0.8113 & & 0.8109 & & 0.8103 & & 0.8115 & & 0.8142 \\
& 3 & 0.8099 & & 0.7870 & & 0.7699 & & 0.7593 & & 0.7536 \\
& 4 & 0.8102 & & 0.7753 & & 0.7486 & & 0.7301 & & 0.7184 \\
& 5 & 0.8097 & & 0.7675 & & 0.7351 & & 0.7118 & & 0.6961 \\
& 6 & 0.8100 & & 0.7622 & & 0.7256 & & 0.6992 & & 0.6808 \\
~~$n$~~ & 7 & 0.8098 & & 0.7582 & & 0.7185 & & 0.6898 & & 0.6696 \\
& 11 & 0.8104 & & 0.7492 & & 0.7021 & & 0.6677 & & 0.6435 \\
& 15 & 0.8110 & & 0.7449 & & 0.6938 & & 0.6565 & & 0.6301 \\
& 21 & 0.8118 & & 0.7416 & & 0.6870 & & 0.6471 & & 0.6188 \\
& 31 & 0.8128 & & 0.7391 & & 0.6814 & & 0.6390 & & 0.6089 \\
\hline\hline
\end{tabular}
\caption{The threshold fidelity $F_\textrm{min}^\textrm{SS}$ as a function of
 $m$ and $n$.
\label{tb:SS_m2-m6}
}
\end{table}

\begin{table}[t]
\begin{tabular}{||c|c|ccccccccc||}
\hline\hline
\multicolumn{2}{||c|}{} & \multicolumn{9}{c||}{$m$} \\
\cline{3-11}
\multicolumn{2}{||c|}{$F_\textrm{min}^\textrm{MS}(m, n)$} &
2 & ~~~~ & 3 & ~~~~ & 4 & ~~~~ & 5 & ~~~~ & 6 \\
\hline
& 2 & 0.8137 & & 0.7788 & & 0.7541 & & 0.7369 & & 0.7253 \\
& 3 & 0.8101 & & 0.7631 & & 0.7261 & & 0.6991 & & 0.6781 \\
& 4 & 0.8102 & & 0.7551 & & 0.7091 & & 0.6781 & & 0.6571 \\
~~$n$~~ & 5 & 0.8095 & & 0.7566 & & 0.7111 & & 0.6771 & & 0.6521 \\
& 6 & 0.8100 & & 0.7522 & & 0.7081 & & 0.6721 & & 0.6421 \\
& 7 & 0.8098 & & 0.7501 & & 0.7051 & & 0.6711 & & 0.6441 \\
& 11 & 0.8104 & & 0.7475 & & 0.6951 & & 0.6581 & & 0.6311 \\
& 15 & 0.8110 & & 0.7446 & & 0.6901 & & 0.6511 & & 0.6221 \\
\hline\hline
\end{tabular}
\caption{The threshold fidelity $F_\textrm{min}^\textrm{MS}$ as a function of
 $m$ and $n$.
\label{tb:MS_m2-m6}
}
\end{table}

\begin{table}[t]
\begin{tabular}{||c|c|ccccccccc||}
\hline\hline
\multicolumn{2}{||c|}{} & \multicolumn{9}{c||}{$m$} \\
\cline{3-11}
\multicolumn{2}{||c|}{$F_\textrm{min}^\textrm{CL}(m, n)$} &
2 & ~~~~ & 3 & ~~~~ & 4 & ~~~~ & 5 & ~~~~ & 6 \\
\hline
& 2 & 0.8137 & & 0.7084 & & 0.6655 & & 0.6378 & & 0.6204 \\
& 3 & 0.8101 & & 0.7122 & & 0.6793 & & 0.6584 & & 0.6501 \\
& 4 & 0.8102 & & 0.7165 & & 0.6906 & & 0.6680 & & 0.6532 \\
~~$n$~~ & 5 & 0.8095 & & 0.7111 & & 0.6776 & & 0.6582 & & 0.6357 \\
& 6 & 0.8100 & & 0.7099 & & 0.6684 & & 0.6551 & & 0.6217 \\
& 7 & 0.8098  & & 0.7086 & & 0.6650 & & 0.6480 & & 0.6133 \\
& 11 & 0.8104 & & 0.7081 & & 0.6642 & & 0.6372 & & 0.6062 \\
& 15 &  0.8110 & & 0.7074 & & 0.6601 & & 0.6284 & & 0.6036 \\
\hline\hline
\end{tabular}
\caption{The threshold fidelity $F_\textrm{min}^\textrm{CL}$ as a function of
 $m$ and $n$.
\label{tb:CL_m2-m6}
}
\end{table}

Whereas for the bipartite case ($m = 2$),
Tables~\ref{tb:SS_m2-m6}--\ref{tb:CL_m2-m6} show that all our three protocols
can tolerate almost the same level of error.  It means that the use of random
asymmetric CSS outer code does not give any significant advantage here.
(Actually, we find that using random asymmetric CSS outer code decreases the
error-tolerant capability for $n \leq 3$.)
Interestingly, the threshold fidelities for our second and third protocols
agree to at least four significant figures for $m = 2$.
This finding can be understood as follows.
As we have discussed in Sec.~\ref{subsec:evaluation} and particularly in
Eq.~(\ref{eq:p_delta_gamma_cond}), the probability of $\delta = 0$ equals
the probability of $\delta = 1$ provided that
$\vec{\boldsymbol{s}} \neq \vec{\boldsymbol{0}}$ irrespective of the value of
$\boldsymbol{\gamma}$.
That is to say, $I(\delta;\gamma|\vec{\boldsymbol{s}}) = 0$ whenever
$\vec{\boldsymbol{s}} \neq \vec{\boldsymbol{0}}$.
So, it is not surprising to find that the weighted mutual information
$\sum_{\vec{\boldsymbol{s}}} \textrm{Pr} (\vec{\boldsymbol{s}})
I(\delta;\boldsymbol{\gamma}| \vec{\boldsymbol{s}})$ becomes negligibly
small when the fidelity $F$ is close to its threshold value.
Combined with Eqs.~(\ref{eq:D_MS})
and~(\ref{eq:D_CL}), it is not unnatural to find that
$F_\textrm{min}^\textrm{MS}(2,n) \approx F_\textrm{min}^\textrm{CL}(2,n)$.

Our numerical computation shows that
$\lim_{n\rightarrow\infty} F_\text{min}^\textrm{SS}(m,n)$,
$\lim_{n\rightarrow\infty} F_\text{min}^\textrm{MS}(m,n)$ and
$\lim_{n\rightarrow\infty} F_\text{min}^\textrm{CL}(m,n)$ are
decreasing functions of $m$.  Besides,
$\lim_{n\rightarrow\infty} F_\text{min}^\textrm{SS}(4,n)$ is smaller
than 0.7798, the fidelity threshold of the Maneva and Smolin's hashing scheme
in the large $m$ limit~\cite{MS00}.  So, for $m\geq 3$, our three protocols
all tolerate a higher noise level than all other one-way schemes known to date.
Figs.~\ref{fig:SS_m2-m5}--\ref{fig:CL_m2-m5} further depict that the yields of
our protocols $D_\textrm{SS}$, $D_\textrm{MS}$ and $D_\textrm{CL}$ are very
steep functions of $F$ around their corresponding threshold fidelities.
Thus, a reasonable yield can be obtained when $F$ is equal to, say,
$0.02$ higher than the threshold.

Another interesting feature found in
Tables~\ref{tb:SS_m2-m6}--\ref{tb:CL_m2-m6} and
Figs.~\ref{fig:SS_m2-m5}--\ref{fig:CL_m2-m5} is that the
threshold fidelities $F_\text{min}^\textrm{SS}(m,n)$,
$F_\textrm{min}^\textrm{MS}(m,n)$ and $F_\textrm{min}^\textrm{CL}(m,n)$ are
all decreasing functions of $n$ for $m \geq 3$. That is, our protocols attain a
higher capacity if players use a longer repetition code whenever $m\geq 3$. In
contrast, DiVincenzo \emph{et al.} found that
$F_\text{min}^\textrm{SS}(2,n)$ attains global minimum when $n = 5$.
Besides, $F_\text{min}^\textrm{SS} (2,n) > F_\text{min}^\textrm{SS}(2,n\pm 1)$
for a small even integer $n$~\cite{DSS98}.
Interestingly, Tables~\ref{tb:MS_m2-m6} and~\ref{tb:CL_m2-m6} show that
$F_\textrm{min}^\textrm{MS}(2,n)$ and $F_\textrm{min}^\textrm{CL}(2,n)$
behave in the same way, too.

Lastly, we remark that
for $m\geq 3$, the improvement in the error-tolerant capability
for increasing $n$ comes with a price.
For a fixed $m\geq 3$, the yields of our protocols decrease as $n$ increases
provided that the channel fidelity $F$ is close to $1$
as depicted in Figs.~\ref{fig:SS_m2-m5}--\ref{fig:CL_m2-m5}.
This is because as $n$
increases, more shared GHZ states must be wasted in order to obtain the error
syndrome $\vec{\boldsymbol{s}}$ even if the channel is noiseless.

\subsection{Understanding the trend of the threshold fidelities}
\label{subsec:explain_trend_q-2}
Although the discussions in this subsection focuses on the trend of the
threshold fidelity of our first protocol, namely, $F_\textrm{min}^\textrm{SS}
(m,n)$, the essential ideas also apply to the cases of our second and third
protocols, that is, $F_\textrm{min}^\textrm{MS}(m,n)$ and
$F_\textrm{min}^\textrm{CL}(m,n)$.

The reason why $F_\text{min}^\textrm{SS}(2,n)$ is a
sawtooth-shaped function of $n$
for $n\lesssim 8$ is related to the behavior of
$h(\{ \text{Pr}((\delta,\boldsymbol{\gamma})|\vec{\boldsymbol{s}}) : (\delta,
\boldsymbol{\gamma}) \in GF(2)^m \})$. It is easy to check that for $m=2$,
$h(\{ \text{Pr}((\delta,\boldsymbol{\gamma})|\vec{\boldsymbol{s}}) : (\delta,
\boldsymbol{\gamma}) \in GF(2)^m \})$ is equal to (much less than) $2$ provided
that the depolarization weight $\text{wt} ( \vec{\boldsymbol{s}}) = n / 2$
($\text{wt} (\vec{\boldsymbol{s}}) \neq n/2$). For a small even $n$, there is
a non-negligible probability of finding $\vec{\boldsymbol{s}}$ with
$\text{wt}(\vec{\boldsymbol{s}}) = n/2$ so that the root of $S_X = 1$ and hence
the value of $F_\text{min}^\textrm{SS}$ are determined mainly by the summing
only over those $\vec{\boldsymbol{s}}$'s with depolarization weight $0$ or
$n/2$ in Eq.~(\ref{eq:Sx_alt}).
In contrast, for a small odd $n$, all entropies in the R.H.S. of
Eq.~(\ref{eq:Sx_alt}) are much less than
$2$. Hence, the corresponding value of $F_\text{min}^\textrm{SS}(2,n)$ is lower
than $F_\text{min}^\textrm{SS}(2,n\pm 1)$. In other words, the reason for
$F_\text{min}^\textrm{SS}(2,n) > F_\text{min}^\textrm{SS}(2,n\pm 1)$ for a
small even $n$ is that there is a non-negligible chance that exactly half of
Bell states used by the inner repetition code
have spin flip error so that players have absolutely no idea what kind
of error the remaining unmeasured Bell state has experienced.

However, the situation is very different when $m\geq 3$. In this case,
the condition for
$h(\{ \text{Pr}((\delta,\boldsymbol{\gamma})|\vec{\boldsymbol{s}}) : (\delta,
\boldsymbol{\gamma}) \in GF(2)^m \}) \geq 2$ is that one can find an integer
$i$ such that $f_{\vec{\boldsymbol{s}}} (i) \geq 2$ and
$f_{\vec{\boldsymbol{s}}}(j) = 0$ for all $j < i$. More importantly, for
a depolarizing channel with $F > 1/2$, the
probability $\text{Pr}(\vec{\boldsymbol{s}})$ of finding this kind of
$\vec{\boldsymbol{s}}$ with
$h(\{ \text{Pr}((\delta,\boldsymbol{\gamma})|\vec{\boldsymbol{s}}) : (\delta,
\boldsymbol{\gamma}) \in GF(2)^m \}) \geq 2$ is much less than that in the
situation of $m = 2$. Thus, the contribution of terms with entropy greater
than or
equal to $2$ in Eq.~(\ref{eq:Sx_alt}) becomes much less significant when $m
\geq 3$.  So, it is not surprising to find that for a fixed $m \geq 3$,
$F_\text{min}^\textrm{SS}(m,n)$ is not a sawtooth-shaped function of $n$ when
$n$ is small.

It is also easy to understand why $\lim_{n\rightarrow\infty}
F_\text{min}^\textrm{SS}
(m,n)$ is a decreasing function of $m$.  One simply check by Taylor's series
expansion that in the limit of large $n$ and for a fixed $1/2 < F < 1$.  Then
we find that the
first term in the R.H.S. of Eq.~(\ref{eq:Sx_alt}) is an increasing 
function of $m$; and that the summand in the second term in the R.H.S.
of Eq.~(\ref{eq:Sx_alt}) is almost surely a decreasing function of $m$ in the
large $n$ limit.

As we have pointed out that the value of $F_\text{min}^\textrm{SS}(m,n)$
depends on the entropy of a few atypical set of errors experienced by the GHZ
states.  We do not have a good explanation why
$F_\text{min}^\textrm{SS}(m,n)$ is a decreasing function of $n$ for $m\geq 3$.

\subsection{Breaking the $F > 0.75$ limit?}
\label{subsec:limit}
No $t$ error correcting quantum code of codeword size $4t$
exists~\cite{KL97a,BDSW96a}.  Hence, it is impossible to
distill Bell states using an one-way scheme provided that the fidelity of the
depolarizing channel is less than or equal to $0.75$~\cite{BDSW96a}.  That is
why $F_\text{min}^\textrm{SS}(2,n), F_\textrm{min}^\textrm{MS}(2,n),
F_\textrm{min}^\textrm{CL}(2,n) > 0.75$.  Interestingly, a few
$F_\text{min}^\textrm{SS}(m,n)$'s,
$F_\text{min}^\textrm{MS}(m,n)$'s and
$F_\text{min}^\textrm{CL}(m,n)$'s listed in
Tables~\ref{tb:SS_m2-m6}--\ref{tb:CL_m2-m6} are less than 0.75.  Does it make
sense?

To solve this paradox, let
us recall that the Pauli errors experienced by a GHZ state shared among $m$
players can always be regarded as taken place in $(m-1)$ of the $m$ qubits.
From Eq.~(\ref{eq:GHZ_gen}), we may regard that at most one of the $(m-1)$
qubits may experience a phase error.  
So, the probability that a depolarized GHZ state has experienced phase error
but not spin flip is $(1-F)/(2^m - 1)$, where $F$ is the channel fidelity.
And the number of erroneous qubits equals $1$ in this case.  Besides,
the probability that exactly $i$ out of the $(m-1)$ qubits
have experienced spin flip is $2(1-F) {m-1 \choose i} / (2^m - 1)$ for
$i=1,2,\ldots , m-1$, where the extra factor of $2$ comes from the fact that
the spin-flipped GHZ state may experience phase shift as well.  Hence, the
average number of erroneous qubits divided by $(m-1)$ is given by
\begin{equation}
 \bar{e} = \frac{1}{m-1} \left[ \frac{1-F}{2^m - 1} +
 \frac{2(1-F)}{2^m - 1} \sum_{i=1}^{m-1} i \enspace {m-1 \choose i} \right]
 = \frac{1-F}{2^m - 1} \left( 2^{m-1} + \frac{1}{m-1} \right) ~.
 \label{eq:average_rate}
\end{equation}
Since no $t$ error correcting quantum code has codeword size less than or
equal to $4t$~\cite{KL97a,BDSW96a}, $\bar{e} < 1/4$.  Consequently, a lower
bound for the threshold fidelities $F_\text{min}^\textrm{x}(m,n)$ (for
$\textrm{x} = \textrm{SS}, \textrm{MS}, \textrm{CL}$) is given by
\begin{equation}
 F_\text{min}^\textrm{x}(m,n) > F_\text{bound} = 1 - \frac{2^m - 1}{4} \left(
 2^{m-1} + \frac{1}{m-1} \right)^{-1} ~. \label{eq:F_min_lower_bound}
\end{equation}
A quick look at the third and the fourth columns in Table~\ref{tb:pw}
convinces us that our protocols do not violate this general limit.
Actually, one of the reasons why we can distill shared GHZ states when $F<0.75$
for $m \geq 3$ is that the average qubit error rate for a depolarized GHZ
state is given by Eq.~(\ref{eq:average_rate}), which is smaller than
$(1-F)$.  Note in particular that in the large $m$ limit, the average qubit
error rate for a depolarized GHZ state is close to $1/2$.  So, it is not
surprising that the bound $F_\text{bound}$ approaches $1/2$ in this case.

\section{Generalization to higher dimensional spin}\label{sec:high_spin}
\subsection{Our extended protocols}
Our three protocols can be generalized to the case when the Hilbert space
dimension of each quantum particle is greater than 2.
That is to say, the $m$ players wanted to share the state
\begin{equation}\label{eq:GHZ_q}
\ket{\Phi^{m+}_q} = \frac{1}{\sqrt{q}}  \sum_{i = 0}^{q-1} \ket{i^{\otimes m}}
\end{equation}
through a depolarizing channel by one-way entanglement distillation.
The quantum codes used in the three generalized protocols are extensions of
their corresponding binary codes to the $q$-nary ones.  In particular, their
common inner code becomes classical $[n,1,n]_q$ repetition code.

We have the following two cases to consider.
\begin{enumerate}
\item For $q = p^m$ where $p$ is a prime number, we may impose a finite
field structure $GF(q)$ to the system by defining
\begin{equation}
X_j: \ket{i} \longmapsto \ket{i + j}
\end{equation}
and
\begin{equation}
Z_j: \ket{i} \longmapsto \omega_p^{\Tr{(ij)}} \ket{i}
\end{equation}
for all $j \in GF(q)$ where $\omega_p$ is a primitive $p$th root of
unity, $\Tr$ is the absolute trace and all arithmetic are performed in the
finite field $GF(q)$.
\item Alternatively, for any integer $q\geq 2$, we may impose a ring
structure ${\mathbb Z}/q{\mathbb Z}$ to the system by defining
\begin{equation}
X_j: \ket{i} \longmapsto \ket{i + j}
\end{equation}
and
\begin{equation}
Z_j: \ket{i} \longmapsto \omega_q^{ij} \ket{i}
\end{equation}
for all $j \in \mathbb{Z}/q \mathbb{Z}$ where $\omega_q$ is a primitive $q$th
root of unity and all arithmetic are performed in the ring
${\mathbb Z}/q{\mathbb Z}$.
\end{enumerate}

From now on, we use the symbol ${\mathbb K}$ to denote either the finite field
$GF(q)$ or the ring ${\mathbb Z}/q {\mathbb Z}$.  Similar to the case of $q =
2$, we use the compact notation $(\beta,\boldsymbol{\alpha}) \equiv (\beta,
\alpha_1, \alpha_2, \ldots, \alpha_{m-1})$ to denote the eigenvalue of the
stabilizer generators where $\beta \in {\mathbb K}$ and
$\boldsymbol{\alpha} \in {\mathbb K}^{m-1}$.

In the qubit case (that is, $q = 2$), the error syndrome measurement is
performed with the aid of CNOT gates.  In the case of $q > 2$, this can be
done via the operator $|i,j\rangle \longrightarrow |i,i-j\rangle$ for all
$i,j\in {\mathbb K}$.  
Suppose the error experienced by the $j$th copy of $\ket{\Phi^{m+}_q}$ is 
$(\beta_j, \boldsymbol{\alpha}_j)$ for $j = 0, \ldots , n - 1$. Then after
measuring the error syndrome for the classical $[n,1,n]_q$ repetition code, we
get $\vec{\boldsymbol{s}} \equiv (\boldsymbol{s}_1, \ldots ,
\boldsymbol{s}_{n-1})$ where
\begin{equation}
\boldsymbol{s}_j \equiv \boldsymbol{\alpha}_j - \boldsymbol{\alpha}_0
\end{equation}
for $1 \leq j \leq n-1$. 
Furthermore, the remaining state shared among the players becomes
\begin{equation}
 (\delta, \boldsymbol{\gamma}) = (\sum_{j = 0}^{n - 1} \beta_j,
 \boldsymbol{\alpha}_0) ~. \label{eq:resultant_state_q}
\end{equation}

\subsection{Finding the capacities of our three generalized protocols}
The analysis in Sec.~\ref{subsec:evaluation} can be easily generalized to the
case of qudits (that is, $q > 3$).  In particular, we prove in the Appendix
that
\begin{equation}
 \text{Pr}((\delta,\boldsymbol{\gamma}) | \vec{\boldsymbol{s}}) = \left\{
 \begin{array}{ll}
  \displaystyle \frac{q^{k-1} (1-F)^k (q^m F - qF + q - 1)^{n-k}}{\sum_i
   f_{\vec{\boldsymbol{s}}} (i) \left[ q (1-F) \right]^i \left( q^m F - q F +
   q - 1 \right)^{n-i}} &
   \enspace \text{if} \enspace k > 0 , \\
  \\
  \displaystyle \frac{( q^m F - q F + q - 1 )^n + (q-1) \left(
   q^m F - 1 \right)^n}{q\sum_i
   f_{\vec{\boldsymbol{s}}} (i) \left[ q (1-F) \right]^i \left( q^m F - q F +
   q - 1 \right)^{n-i}} & \enspace \text{if} \enspace k = 0 \enspace \text{and}
   \enspace \delta = 0 , \\
  \\
  \displaystyle \frac{( q^m F - q F + q - 1 )^n - \left( q^m F - 1
   \right)^n}{q\sum_i
   f_{\vec{\boldsymbol{s}}} (i) \left[ q (1-F) \right]^i \left( q^m F - q F +
   q - 1 \right)^{n-i}} & \enspace \text{if} \enspace k = 0 \enspace \text{and}
   \enspace \delta \neq 0 ,
 \end{array}
 \right. \label{eq:qudit_pr}
\end{equation}
where $k$ and $f_{\vec{\boldsymbol{s}}}(i)$ are given by
Eq.~(\ref{eq:k_def}) and Eq.~(\ref{eq:f_s_qudit}) respectively.

The yields of our three generalized protocols can be computed using
Eqs.~(\ref{eq:D_SS})--(\ref{eq:Sx}) and~(\ref{eq:D_MS})--(\ref{eq:D_hDef3})
just like the case of $q=2$.  Nevertheless, there is an important subtlety.
Since the players can make full use of each of the $q$ possible error syndrome
measurement outcomes to distill the generalized GHZ state
$|\Phi_q^{m+}\rangle$, the entropies and conditional entropies in
Eqs.~(\ref{eq:Sx}), (\ref{eq:D_hDef1}), (\ref{eq:D_hDef2})
and~(\ref{eq:D_hDef3}) should be measured in the unit of dit rather than bit.
That is to say, the base of the logarithm used in these entropies should be
$q$ instead of $2$.
And since the dimension of each information carrier $q$ has
changed, one should not directly compare the yields of the qudit-based
protocols with those of the standard qubit-based ones.
Note further that similar to the original qubit-based protocols, we can compute
these yields in a time sub-exponential in $n$.

\subsection{Performance of our generalized protocols}
\begin{figure}[t]
\centering\includegraphics[width=14.5cm]{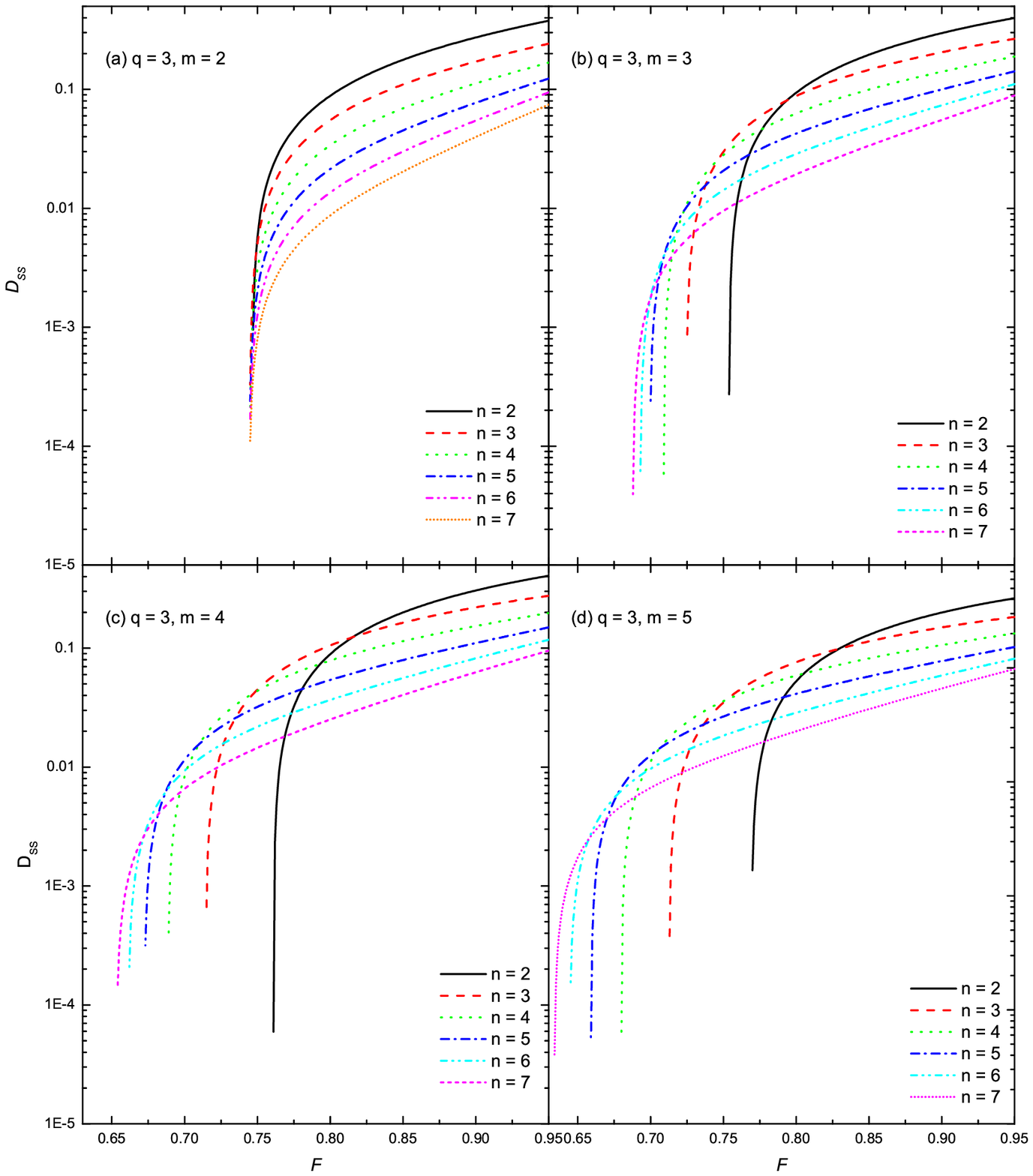}
 \caption{The yield $D_\textrm{SS}$ of our first generalized protocol for
  distilling $\ket{\Phi^{m+}_3}$ after passing through a depolarizing channel
  of fidelity $F$ using the classical repetition code $[n, 1, n]_q$ as the
  inner code for various $n$ when (a)~$m = 2$, (b)~$m = 3$, (c)~$m = 4$ and
  (d)~$m = 5$.
\label{fig:SS_q-3-m2-m5}
}
\end{figure}

\begin{figure}[t]
\centering\includegraphics[width=14.5cm]{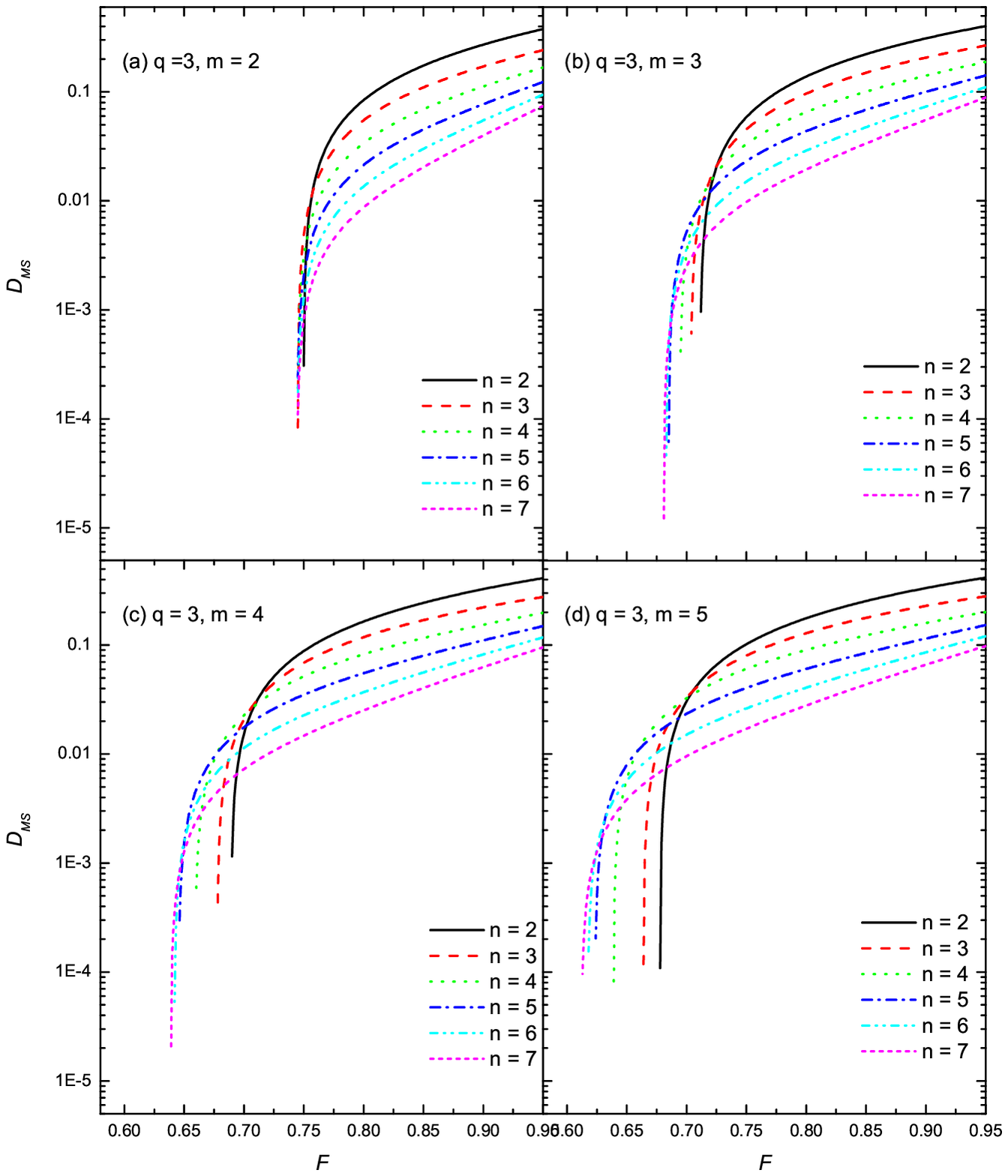}
 \caption{The yield $D_\textrm{MS}$ of our second generalized protocol for
  distilling $\ket{\Phi^{m+}_3}$.  The parameters used are the same as those
  in Fig.~\ref{fig:SS_q-3-m2-m5}.
\label{fig:MS_q-3-m2-m5}
}
\end{figure}

\begin{figure}[t]
\centering\includegraphics[width=14.5cm]{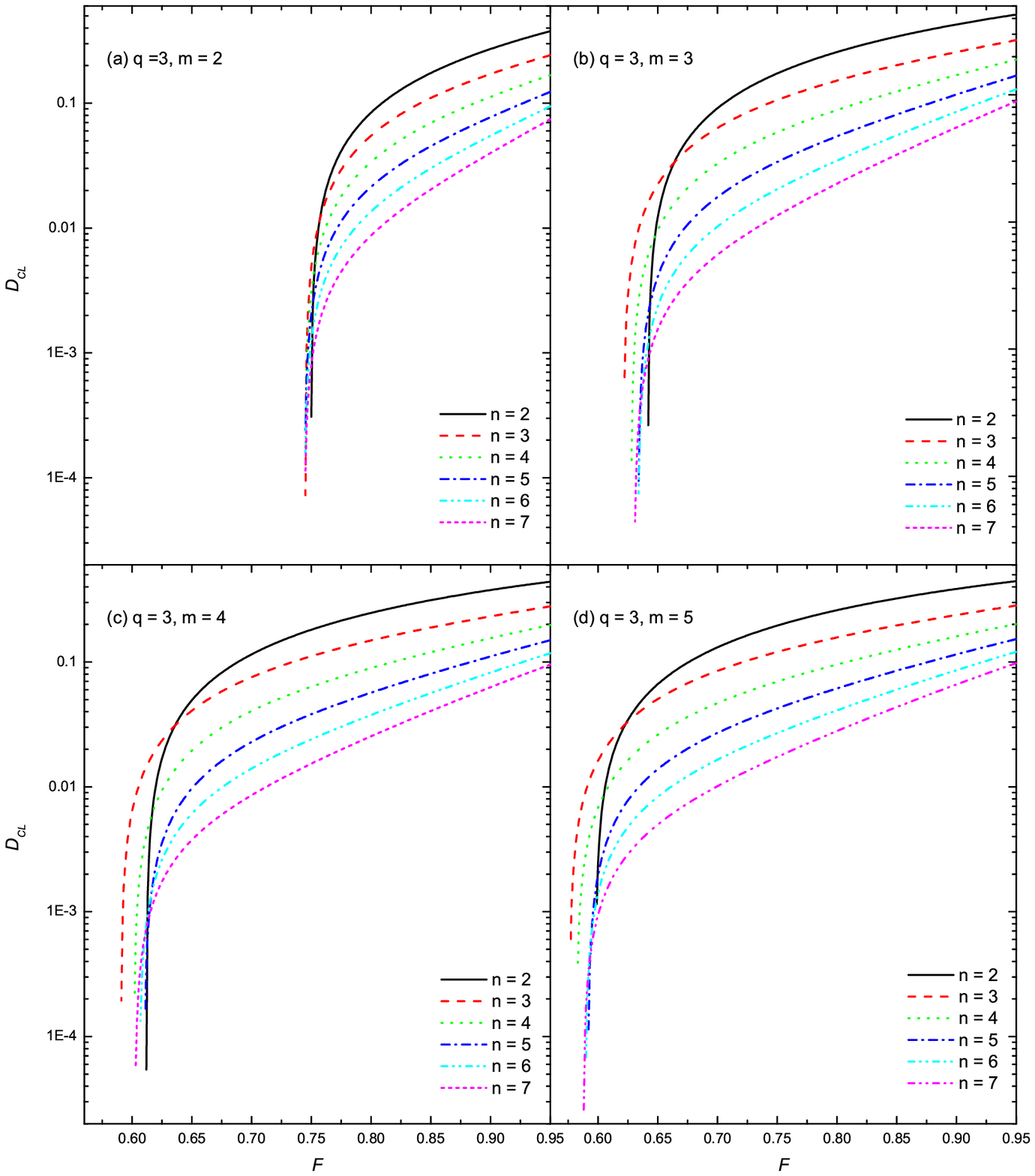}
 \caption{The yield $D_\textrm{CL}$ of our third generalized protocol for
  distilling $\ket{\Phi^{m+}_3}$.  The parameters used are the same as those
  in Fig.~\ref{fig:SS_q-3-m2-m5}.
\label{fig:CL_q-3-m2-m5}
}
\end{figure}

Figs.~\ref{fig:SS_q-3-m2-m5}--\ref{fig:CL_q-3-m2-m5} depict the yields of our
three generalized protocols in the case of $q = 3$.
And Tables~\ref{tb:SS_q-3-m2-m6}--\ref{tb:CL_q-3-m2-m6} list the
corresponding threshold fidelities.  Clearly, the trend of the threshold
fidelities of our three generalized protocols are very similar to their
corresponding cases of $q = 2$.
In particular, for $q = 3$, the threshold fidelities
$F_\textrm{min}^\textrm{SS} (m,n)$, $F_\textrm{min}^\textrm{MS} (m,n)$ and
$F_\textrm{min}^\textrm{CL} (m,n)$ are decreasing function of $n$ for any
fixed integer $m\geq 3$; while they reach global minima at $n = 7$ provided
that $m = 2$.
These findings are not completely surprising because the arguments we have
used to explain the trends of the yields and threshold fidelities for our
three protocols in the case of $q = 2$ reported in
Sec.~\ref{subsec:explain_trend_q-2} are also applicable here after minor
adjustments.

\begin{table}[t]
\begin{tabular}{||c|c|ccccccccc||}
\hline\hline
\multicolumn{2}{||c|}{} & \multicolumn{9}{c||}{$m$} \\
\cline{3-11}
\multicolumn{2}{||c|}{$F_\textrm{min}(m, n)$} &
2 & ~~~~ & 3 & ~~~~ & 4 & ~~~~ & 5 & ~~~~ & 6\\
\hline
& 2 & 0.7462 & & 0.7538 & & 0.7609 & & 0.7693 & & 0.7778 \\
& 3 & 0.7445 & & 0.7243 & &0.7145 & & 0.7127 & & 0.7148  \\
& 4 & 0.7445 & &  0.7089& & 0.6885 & & 0.6799 & & 0.6779 \\
~~$n$~~ & 5 & 0.7442 & & 0.6994 & & 0.6722 & & 0.6588 & & 0.6539 \\
& 6 & 0.7441 & & 0.6927 & & 0.6611 & & 0.6443 & & 0.6370 \\
& 7 & 0.7441  & & 0.6877 & & 0.6530 & & 0.6337 & & 0.6246 \\
& 11 & 0.7444 & & 0.6759 & & 0.6338 & & 0.6096 &&  0.5962 \\
& 15 & 0.7449 & & 0.6700 & & 0.6238 & & 0.5974 & & 0.5823 \\
\hline\hline
\end{tabular}
\caption{The threshold fidelity $F_\textrm{min}^\textrm{SS}$ as a function of
 $m$ and $n$ for $q = 3$.
\label{tb:SS_q-3-m2-m6}
}
\end{table}

\begin{table}[t]
\begin{tabular}{||c|c|ccccccccc||}
\hline\hline
\multicolumn{2}{||c|}{} & \multicolumn{9}{c||}{$m$} \\
\cline{3-11}
\multicolumn{2}{||c|}{$F_\textrm{min}^\textrm{MS}(m,n)$} &
2 & ~~~~ & 3 & ~~~~ & 4 & ~~~~ & 5 & ~~~~ & 6 \\
\hline
& 2 & 0.7499 & & 0.7114 & & 0.6892 & & 0.6780 & & 0.6728 \\
& 3 & 0.7450 & & 0.7034 & & 0.6776 & & 0.6640 & & 0.6575 \\
& 4 & 0.7448 & & 0.6944 & & 0.6591 & & 0.6389 & & 0.6289 \\
~~$n$~~ & 5 & 0.7444 & & 0.6849 & & 0.6452 & & 0.6234 & & 0.6127 \\
& 6 & 0.7443 & & 0.6829 & & 0.6418& & 0.6172 & & 0.6041 \\
& 7 & 0.7443 & & 0.6810& & 0.6390 & & 0.6121 & & 0.5967 \\
& 11 & 0.7446 & & 0.6738 & & 0.6289& & 0.5997 & & 0.5808 \\
& 15 & 0.7451 & & 0.6692& & 0.6212 & & 0.5927 & & 0.5740 \\
\hline\hline
\end{tabular}
\caption{The threshold fidelity $F_\textrm{min}^\textrm{MS}$ as a function of
 $m$ and $n$ for $q = 3$.
\label{tb:MS_q-3-m2-m6}
}
\end{table}

\begin{table}[t]
\begin{tabular}{||c|c|ccccccccc||}
\hline\hline
\multicolumn{2}{||c|}{} & \multicolumn{9}{c||}{$m$} \\
\cline{3-11}
\multicolumn{2}{||c|}{$F_\textrm{min}^\textrm{CL}(m,n)$} &
2 & ~~~~ & 3 & ~~~~ & 4 & ~~~~ & 5 & ~~~~ & 6 \\
\hline
& 2 & 0.7499  & & 0.6419& & 0.6120 & & 0.5981 & & 0.5921 \\
& 3 & 0.7450 & & 0.6222& & 0.5908 & & 0.5762 & & 0.5697 \\
& 4 & 0.7448 & & 0.6282 & & 0.6016& & 0.5821 & & 0.5868 \\
~~$n$~~ & 5 & 0.7444 & & 0.6337 & & 0.6104 & & 0.5916 & & 0.5895 \\
& 6 & 0.7443 & & 0.6334 & & 0.6061& & 0.5896 & & 0.5855 \\
& 7 & 0.7443 & & 0.6304& & 0.6023 & & 0.5877 & & 0.5829 \\
& 11 & 0.7446 & & 0.6249 & & 0.5910& & 0.5790 & & 0.5670 \\
& 15 & 0.7451 & & 0.6235& & 0.5865 & & 0.5701 & & 0.5560 \\
\hline\hline
\end{tabular}
\caption{The threshold fidelity $F_\textrm{min}^\textrm{CL}$ as a function of
 $m$ and $n$ for $q = 3$.
\label{tb:CL_q-3-m2-m6}
}
\end{table}

\subsection{Lower bound for the three threshold fidelities}
The proof that no $t$ error correcting quantum code with codeword size $4t$
is also applicable to qudits~\cite{R99}.  We may use this fact to establish a
lower bound for the threshold fidelities of our three generalized protocols
when qudits are used as information carriers.  Since the proof is also the same
as that of the qubit case reported in Sec.~\ref{subsec:limit}, here we only
write down the bound without giving the details of the proof:
\begin{equation}
 F_\text{min}^\textrm{SS} (m,n), F_\text{min}^\textrm{MS} (m,n),
 F_\text{min}^\textrm{CL} (m,n) > F_\text{bound} = 1 - \frac{q^m - 1}{4(q-1)}
 \left( q^{m-1} + \frac{1}{m-1} \right)^{-1} ~.
\end{equation}

\section{Summary and Discussion}\label{sec:sum}
In summary, we have introduced three one-way GHZ state purification protocols
using degenerate codes by extending the works of DiVincenzo
\emph{et al.} on one-way Bell state purification via degenerate
codes~\cite{SS96,DSS98} as well as the works of Maneva and Shor~\cite{MS00}
and its generalization by Chen and Lo~\cite{CL07} on multipartite entanglement
purification using random asymmetric CSS codes.
Then, we calculate the yields of our three protocols when the inputs are Werner
states.  The method
we used to calculate these yields is divided into two steps.  The first step is
to calculate entropies or conditional entropies such as
$h(\{ \textrm{Pr}((\delta,\boldsymbol{\gamma}) : (\delta,\boldsymbol{\gamma})
\in GF(2)^m \})$ by means of the so-called depolarization weight
enumerator.  Actually, the first step can be easily extended to the case
of an arbitrary stabilizer inner code, an arbitrary un-correlated noise
model and an arbitrary stabilizer output state.
The second step involves the computation of a weighted sum of the
entropies or conditional entropies obtained in the first step.
Nonetheless, for a general stabilizer inner code, a general un-correlated
noise model and a general output state, this sum may not be practical as it
involves up to about $2^{m n}$ number of terms.
Fortunately, as the inner code used in our purification scheme is the highly
symmetrical classical repetition code,
we are able to greatly simplify the sum, making the computation of the
yields in a time which is sub-exponential in $n$ when the GHZ states are
subjected to depolarization errors.
In this way, we can calculate the corresponding threshold fidelities
accurately and reasonably fast.  This is quite an accomplishment because
finding the threshold fidelities involves the accurate determination of the
sign of the difference between two small positive numbers provided that the
number of players $m\geq 3$ and the codeword size of the inner repetition code
$n$ is large.  (See, for example, Eq.~(\ref{eq:Sx_alt}).)
Just like the Bell state case, we discover that the threshold fidelities
of our three protocols are better than all known one-way GHZ state
purification schemes to date. So, once again, the power of using degenerate
codes to combat quantum errors is demonstrated.

We also extended our scheme to tackle the case when the information carriers
are qudits instead of qubits.  We find that the performance trend of these
generalized schemes are quite similar to those of the qubit cases.

There are a few un-answered questions, however.  Here we list some of them.
The reason why the threshold fidelities $F_\text{min}^\textrm{SS}$
$F_\textrm{min}^\textrm{MS}$ and $F_\textrm{min}^\textrm{CL}$ decrease with
$n$ for $m\geq 3$ is not apparent.
And apart from the general statement that degenerate codes pack more
information than non-degenerate ones making them powerful in one-way
purification of GHZ states, can we specifically understand why using classical
repetition code concatenated with a random hashing quantum code is more
error-tolerant than a few other choices of degenerate codes?~\cite{HC08}
Along a different line, it is important to find out the value of
$\lim_{m\rightarrow\infty} \lim_{n\rightarrow\infty} F_\text{min}^\textrm{CL}
(m,n)$ and compare it with the $1/2$ lower bound.
Finally, it is instructive to extend our study to the case of using a different
degenerate code to distill another type of entangled state subjected to
another noise model, such as the Pauli channel~\cite{SS06}.

\acknowledgments
Useful discussions with C.-H.~F. Fung and H.-K. Lo are gratefully acknowledged.
This work is supported by the RGC grants No.~HKU~7010/04P and No.~HKU~701007P
of the HKSAR Government.

\appendix
\section{Proof Of Eq.~(\ref{eq:qudit_pr})}
We prove the validity of Eq.~(\ref{eq:qudit_pr}) by following the analysis in
Sec.~\ref{subsec:evaluation}. (And we follow the same notations as used in
Sec.~\ref{subsec:evaluation} after possibly some straight-forward extension
to the case of qudits.)
First, we extend the definition of depolarization weight as follows. Let
$(\beta_j,\boldsymbol{\alpha}_j)_{j=0}^{n-1} \in {\mathbb K}^{m n}$
be a ordered $n$-tuple.  Then its depolarization weight is
defined as the Hamming weight by regarding this $n$-tuple as a vector
of elements in ${\mathbb K}$.
Clearly, $\text{Pr}((\delta,\boldsymbol{\gamma})\wedge\vec{\boldsymbol{s}}) =
\text{Pr}({\mathfrak E}(\vec{\boldsymbol{s}},\delta,\boldsymbol{\gamma}))$
where
\begin{equation}
{\mathfrak E}(\vec{\boldsymbol{s}},\delta,\boldsymbol{\gamma}) = \{
(\beta_j,\boldsymbol{\alpha}_j)_{j=0}^{n-1} \in {\mathbb K}^{m n} : \delta =
\sum_{j=0}^{n-1} \beta_j , \boldsymbol{\gamma} = \boldsymbol{\alpha}_0,
\boldsymbol{\alpha}_\ell = \boldsymbol{s}_\ell + \boldsymbol{\alpha}_0
\enspace \text{for} \enspace \ell = 1,2,\ldots,n-1 \} ~.
\end{equation}

Using the same argument as in the case of $q = 2$, we conclude that for
$k > 0$,
\begin{eqnarray}
w({\mathfrak E}(\vec{\boldsymbol{s}},\delta,\boldsymbol{\gamma});x,y) &=& 
\frac{1}{q} \sum_{ \{ a_{i, \boldsymbol{t}} \} }
\frac{k!}{\displaystyle \prod_{\genfrac{}{}{0pt}{}{i\in {\mathbb K},}{\boldsymbol{t}
\in {{\mathbb K}^{m-1}}\setminus \{ \boldsymbol{0} \}}} a_{i,\boldsymbol{t}}!}
\enspace
\frac{(n-k)!}{\displaystyle \prod_{i\in {\mathbb K}^*} a_{i,\boldsymbol{0}}!}
\enspace x^{n -a_{0, \boldsymbol{0}}} y^{a_{0,\boldsymbol{0}}}
\nonumber \\
&=& q^{k -1} \sum_{a_{0, \boldsymbol{0}}}
{n - k \choose a_{0, \boldsymbol{0}}} \enspace (q-1)^{n-k-a_{0,\boldsymbol{0}}}
x^{n - a_{0, \boldsymbol{0}}} y^{a_{0,\boldsymbol{0}}} \nonumber \\
&=& q^{k-1} x^k [(q-1)x+y]^{n-k} ~.
\end{eqnarray}
For $k = 0$, we have to use a slightly different method to compute the
depolarization weight enumerator.
By substituting $x_0 = y$ and $x_\eta = x$ for all $\eta\neq 0$ into the
identity
\begin{equation}
 \left( \sum_{\eta\in {\mathbb K}} x_\eta \right)^n = \sum_{\{ a_\eta \}}
 \left( \frac{n!}{\prod_{\eta\in {\mathbb K}} a_\eta!} \prod_{\eta\in
 {\mathbb K}} x_\eta^{a_\eta} \right) ~, \label{eq:multisum}
\end{equation}
we have
\begin{equation}
 \sum_{\delta\in{\mathbb K}} w({\mathfrak E}(\vec{\boldsymbol{s}},\delta,
 \boldsymbol{\gamma});x,y) = [ (q-1)x + y ]^n ~.
\end{equation}
If ${\mathbb K} = GF(q)$, then by putting $x_0 = y$ and
$x_\eta = \omega_p^{\Tr (\eta \rho)} x$ for all $\eta\neq 0$ where $\rho\in
GF(q)^*$ into Eq.~(\ref{eq:multisum}), we have
\begin{equation}
 \sum_{\delta\in GF(q)} \omega_p^{\Tr (\delta \rho)}
 w({\mathfrak E}(\vec{\boldsymbol{s}},\delta,\boldsymbol{\gamma});x,y) =
 (y-x)^n ~.
\end{equation}
If ${\mathbb K} = {\mathbb Z}/q {\mathbb Z}$, then we put $x_0 = y$
and $x_\eta = \omega_q^{\eta\rho)} x$ for all $\eta\neq 0$ where $\rho\in
({\mathbb Z}/q {\mathbb Z})^*$ into Eq.~(\ref{eq:multisum}), we
arrive at
\begin{equation}
 \sum_{\delta\in {\mathbb Z}/q {\mathbb Z}} \omega_q^{\delta \rho}
 w({\mathfrak E}(\vec{\boldsymbol{s}},\delta,\boldsymbol{\gamma});x,y) =
 (y-x)^n ~.
\end{equation}
Consequently, we conclude that for ${\mathbb K} = GF(q)$ or ${\mathbb Z}/q
{\mathbb Z}$,
\begin{equation}
w({\mathfrak E}(\vec{\boldsymbol{s}},\delta,\boldsymbol{\gamma});x,y) =
\left\{ \begin{array}{ll} 
\displaystyle q^{k - 1} x^k \left[ (q-1)x+y \right]^{n-k} & \enspace
 \text{if} \enspace k > 0 ,
\\
\\ \displaystyle
\frac{1}{q} \left\{ \left[ (q-1) x + y \right]^n + (q-1) ( y - x)^n \right\} &
\enspace \text{if} \enspace k = 0
\enspace \text{and}\enspace \delta = 0, \\
\\
\displaystyle
\frac{1}{q} \left\{ \left[ (q-1) x + y \right]^n - (y-x)^n \right\} &
\enspace \text{if} \enspace k = 0 \enspace \text{and}\enspace
\delta \neq 0 . \end{array} \right.
\label{eq:weight_qudit}
\end{equation}

Surely, for depolarizing channel,
$\text{Pr}((\delta,\boldsymbol{\gamma}) \wedge \vec{\boldsymbol{s}}) =
w({\mathfrak E}(\vec{\boldsymbol{s}},\delta,\boldsymbol{\gamma});
(1-F)/(q^m - 1),F)$ and
\begin{eqnarray}
 \text{Pr}(\vec{\boldsymbol{s}}) & = & \sum_{\boldsymbol{t}} \left[
 \frac{q(1-F)}{q^m - 1} \right]^{k(\vec{\boldsymbol{s}},\boldsymbol{t})} \left[
 F + \frac{(q-1)(1-F)}{q^m - 1} \right]^{n-k(\vec{\boldsymbol{s}},
 \boldsymbol{t})} \nonumber \\
 & = & \frac{1}{(q^m - 1)^n} \sum_{i=0}^n f_{\vec{\boldsymbol{s}}} (i)
 \left[ q (1-F) \right]^i \left(
 q^m F - q F + q - 1 \right)^{n-i} ~, \label{eq:prs_qudit}
\end{eqnarray}
where
\begin{equation}
f_{\vec{\boldsymbol{s}}} (i) = |\{ \boldsymbol{t} \in {\mathbb K}^{m - 1} :
k(\vec{\boldsymbol{s}},\boldsymbol{t}) = i \} | ~.
\label{eq:f_s_qudit}
\end{equation}
Combining Eqs.~(\ref{eq:weight_qudit})--(\ref{eq:f_s_qudit}), we arrive at
Eq.~(\ref{eq:qudit_pr}).
\hfill$\Box$

\bibliography{qc39.2}
\end{document}